# Droplet impact and Leidenfrost dynamics on a heated post

**Junhui Li[1], Patricia Weisensee[1,2]**

[1]*Mechanical Engineering and Materials Science, Washington University in St. Louis, St. Louis, MO 63130*

[2]*The Institute of Materials Science & Engineering, Washington University in St. Louis, St. Louis, MO 63130*

Highlights

- A millimetric post raises the Leidenfrost point by 20°C compared to a flat surface
- The post promotes two modes of temperature-dependent droplet breakup
- For nucleate boiling, the post increases the droplet cooling capacity by up to 24%
- When tilted, the post prevents droplet sliding and rebound, increasing heat transfer

Abstract

This study experimentally explores fluid breakup and Leidenfrost dynamics for droplets impacting a heated millimetric post. Using high-speed optical and infrared imaging, we investigate the droplet lifetime, breakup and boiling modes, as well as the cooling performance of different substrates. The post substrate leads to a shorter droplet lifetime and a 20 °C higher Leidenfrost temperature compared to a flat substrate, attributed to mixed boiling modes along the height of the post and additional pinning. For temperatures below the Leidenfrost point, in the nucleate boiling regime, the post substrate also provides a larger maximum temperature drop than its flat counterpart. The enhanced cooling capacity can be attributed to better droplet pinning and an enlarged droplet-substrate contact area. The post's superior cooling performance becomes especially clear for impact on an inclined surface, where the post successfully prevents the rolling and bouncing of the droplet, providing a 51% to 180% increase in the maximum local temperature drop. Interestingly, at temperatures slightly above the Leidenfrost point and for a relatively narrow range of Weber numbers, droplets on the post substrate rebound rather than break up due to a complex interplay of inertial, capillary, and bubble (thin film) dynamics. Overall, the findings show that a large-scale structure increases the Leidenfrost temperature and enhances droplet breakup and interfacial heat transfer efficiency during non-isothermal droplet impact.

# 1 Introduction

Research on droplets impacting heated surfaces is motivated by various applications such as spray cooling [1, 2] and internal combustion engines [3, 4]. During such non-isothermal droplet impact, the hydrodynamics and heat transfer are strongly influenced by the surface temperature, and vice versa [5]. At the lowest temperatures, droplet dynamics and phase change are mostly decoupled, with contact-line-dominated evaporation taking place after droplet deposition [6, 7]. When the surface temperature is higher than the liquid saturation point, the droplet undergoes nucleate boiling and the phase change can have a pronounced effect on droplet dynamics and heat transfer characteristics [8]. Surface cooling is most efficient in this temperature range. Ultimately, if the surface temperature is higher than the Leidenfrost temperature, $T_L$, the droplet remains separated from the hot solid surface by a developing vapor layer, which acts as a thermal barrier and significantly reduces the heat transfer [9, 10]. The droplet dynamics and $T_L$ are influenced by a number of parameters, including wettability [11, 12], surface roughness [13], solid surface thermal conductivity [14], ambient pressure [15, 16], and types of fluids [17]. For example, for droplet impact on a hydrophilic surface, $T_L$ is increased when compared to a non-wetting counter-part, since the viscous dissipation from a larger contact area requires higher superheat to trigger rebounding dynamics and a stable film boiling state [11]. On the other hand, $T_L$ and the cooling capacity decrease for substrates with a higher thermal conductivity, due to a higher interface temperature that increase the vapor thickness [14].

The substrate morphology also strongly contributes to the behavior of the impacting droplet [18]. Micro and nanostructured surfaces have been widely studied to explore their influence on heat transfer and $T_L$ [13, 19-23]. The static Leidenfrost temperature, *i.e.*, $T_L$ of a droplet gently deposited on a surface, increases on microstructure-array surfaces with sparse structure spacings, as the pressure in the vapor decreases thanks to the gaps in the structures and becomes insufficient to support the liquid droplet [19, 20]. Quantitatively, the static Leidenfrost temperature is found to increase with the effective permeability of the surface, which is a function of the micropillar spacing and height and the excess vapor gap [21]. Other micro/nanostructured surfaces influence the droplet behavior through enhanced wettability and wicking, which prevent the droplet from detaching from the surface. Kruse *et al.* [22] used a micro/nanostructured surface fabricated by femtosecond laser processing and increased $T_L$ by up to 175 °C, which was attributed to the reduced contact angle and substantial capillary wicking. A similar observation was also reported on multiscale micro/nano-textured zirconium surfaces [24]. Silicon surfaces covered with carbon-nanofibers delay the transition to film boiling to 200°C higher temperatures compared to smooth surfaces for FC-72 droplets [25]. Combining small-scale structures with other surface designs, a structured surface that consist of small steel pillars, an embedded insulating membrane, and U-shaped channels was created that had a Leidenfrost temperature of 1150°C - a record thus far [26]. In addition to influencing the static Leidenfrost temperature, structured surfaces can also have an effect on the dynamic Leidenfrost temperature, which is defined as the temperature at which the droplet will not wet the surface during impact, *i.e.*, when the droplet has a non-negligible impact velocity [27]. At a given interspacing and pillar width, the dynamic Leidenfrost temperature is found to increase with Weber number, but to decrease with increasing pillar height [28]. At the same Weber number, the dynamic $T_L$ on textured surfaces decreases as the pillar pitch becomes finer, and can even be lower than that of the smooth surface [23].

While small-scale structures influence the impact dynamics through a modification of the vapor dynamics beneath the droplet, a structure with a similar dimension to the droplet induces a complete deformation of the entire droplet shape and thus significantly changes the droplet impact behavior [29]. The pining force caused by sharp edges and the additional contact area between the structure and the liquid also influences the droplet profile [30, 31]. Several studies investigated the hydrodynamics for droplet impact on solid stand-alone structures at isothermal conditions, where heat transfer is neglected [29, 32-34]. For example, Ding *et al.* [35] studied the droplet impact dynamics on single-post superhydrophobic surfaces and found that the rebound and breakup morphologies are dependent on impact Weber number and the ratio between droplet and post diameters. For droplet impact on a cubic pillar, air entrapment around the pillar side faces occurs and varies with the exact impact location on the pillar [33].

In the case of a heated substrate, one would expect this gas entrapment and the altered rebound and breakup dynamics to have a significant influence on the heat transfer between droplet and substrate by serving as nucleation sites for vapor bubbles, preventing or promoting droplet departure, and influencing the Leidenfrost temperature. Furthermore, for impact on a millimetric cylindrical post, the relatively large structure is expected to not only increase the liquid-solid interfacial area, but also decrease the conduction resistance that the bulk liquid usually poses after droplet deposition, effectively enhancing heat transfer rates. Additional advantages of using millimetric structures made of metal as compared to silicon or nanoparticle-based microstructures are their superior durability, ease of manufacturing, and seamless integration with existing metallic components. However, to the best of the authors' knowledge, the combined effect of droplet impact dynamics, heat transfer, and Leidenfrost dynamics on a heated large-scale structure remains unknown. In this study, we thus investigated non-isothermal droplet impact and Leidenfrost dynamics on a single post structure on an aluminum surface, which is easily fabricated using traditional machining methods, and compared them to impact on a flat surface. The droplet lifetime and the static Leidenfrost temperature were analyzed. For droplet impact, typical droplet impact and boiling modes, as well as the breakup morphologies, were observed and classified for Weber numbers ranging from 10 to 120 and surface temperatures between 120 °C and 330 °C, which allows us to infer dynamic Leidenfrost temperatures as well. The cooling effect of the droplets on the substrates was also quantified by measuring the transient temperature distribution at the bottom of the thin substrates using high-speed infrared (IR) imaging. Such direct temperature measurements provide a better understanding of the interplay of hydrodynamics and cooling and allow thermal scientists and engineers to design superior thermal management applications. In addition, droplet dynamics and heat transfer characteristics on an inclined surface were compared to highlight the important contribution of pinning at the post structure on enhancing the cooling efficiency.

## 2 Experimental method

### 2.1 Sample preparation

In this study, two types of aluminum substrates were used: a flat substrate and a (single) post substrate. Both substrates were machined using a Bantam Desktop CNC Milling Machine with a flat-end mill and have a thickness of 1.1 mm (± 0.03 mm). The two substrates share the same surface roughness $R_a = 0.5$ µm and $R_z = 2.1$ µm, as measured by a profilometer (KLA-Tencor Alpha-Step D-100). For the post substrate, a cylindrical structure (1 mm ± 0.03 mm in height and 1 mm ± 0.06 mm in diameter) was machined at the center of the substrate. Before use, substrates were flushed with acetone, isopropyl alcohol (IPA) and de-ionized (DI) water to remove residues from the manufacgturing process. However, even short-term exposure of a clean hydrophilic solid to the open atmosphere can cause accumulation of hydrophobic organic contaminants [36]. Consequently, the equilibrium contact angle of water on the here-used aluminum substrates was $\theta_{eq} = 82° ± 2°$ at room temperature. The advancing and receding contact angles were measured to be $\theta_{adv} = 108° ± 2°$, and $\theta_{rec} = 23° ± 2°$, respectively. The bottom (back side) of the substrates were spray-coated with a ≈5 µm thin layer of black paint (Krylon Latex Enamel), which serves as a transducer for infrared (IR) temperature measurements. No other coatings, which are often prone to degradation [37, 38], were applied, as to best represent possible industrial applications, such as spray cooling for metal quenching [39] and in-wheel motor thermal management in electric vehicles [40].

### 2.2 Experimental setup

A schematic of the experimental setup is shown in Fig. 1. The aluminum substrates were placed on two heating blocks made from copper, which were embedded with cartridge heaters (1/8" Diameter, 120V, 10W) to provide stable heating power to the substrate. Thermal insulation was applied surrounding the heated blocks to reduce heat losses and to keep the temperature stable. The side-view shadowgraph images of the impacting droplets were recorded at 1000 to 5000 frames per second (fps) using a high-speed camera (Photron Mini AX200, Japan) with a Canon MP-E 65 mm f/2.8 1-5X Macro Lens (Japan) at a spatial resolution of 3.3 µm/pixel. A high-speed mid-wave IR camera (Telops FAST M3k, Canada), equipped with a 1x long working distance lens (Telops, Canada), recorded the thermal signals of the substrates in bottom view at 100 fps to 500 fps with a spatial resolution of 30 µm/pixel. Note that we measured the temperature distribution of the black paint layer (≈ substrate bottom surface temperature) and assumed a uniform temperature through the sample prior to impact due to the high thermal conductivity and small thickness of the aluminum substrates. In reality, there is a small lag in the thermal profile at the bottom of the substrate compared to the droplet-substrate interface ($t \cong \frac{1}{\alpha L^2} \approx 0.1\ s$, where $\alpha$ is the thermal diffusivity of the sample and $L$ is its thickness), which is not relevant for the purpose of this study. Droplets of DI water were generated at the tip of a needle (gauge 25) connected to a syringe pump (New Era NE-1000, United States). Droplets with a diameter of 2.7 ± 0.05 mm (droplet volume ≈10.3 µL) detached due to gravity from needles mounted at heights between 3 mm and 200 mm, leading to impact velocities ranging from 0.15 m/s to 1.8 m/s, as determined from the analysis of side-view high-speed sequences of droplets just prior to impact. The corresponding Weber numbers, defined as $We = \rho \frac{Dv^2}{\sigma}$, where $\rho$, $\sigma$ are the density and surface tension of water at room temperature, $D$ is the droplet diameter, and $v$ is the droplet velocity right before impacting the surface, are between 1.5 and 120.

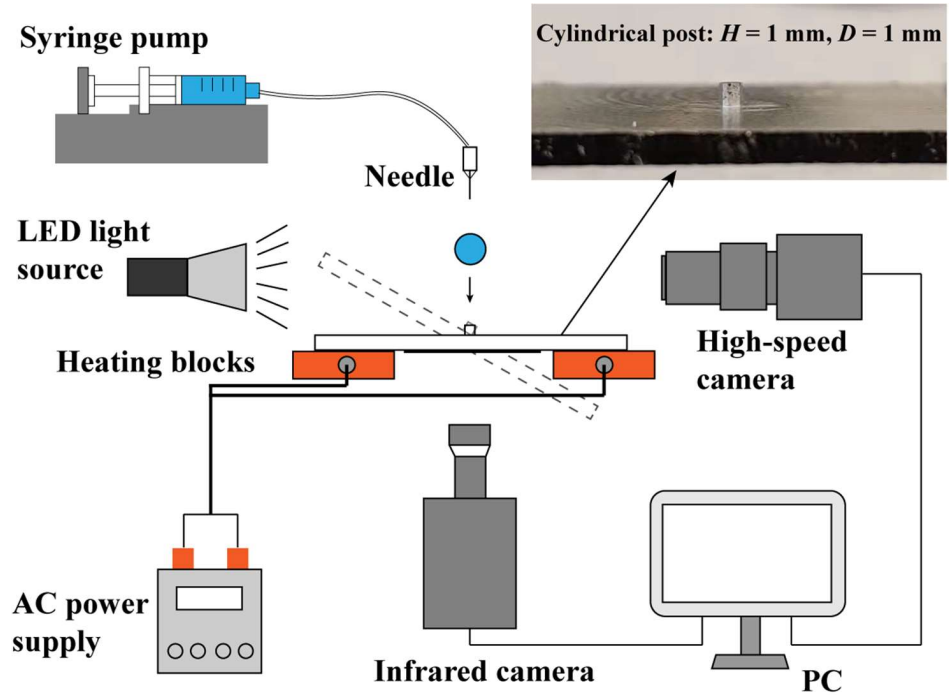

Fig. 1 Schematic of the experimental setup. The image of the post substrate is provided in the top right corner. The droplet impact experiments are conducted for horizontal and inclined (shown as the dashed line) substrates. For experiments on an inclined surface, the IR camera also tilted by the same angle (not shown here).

## 2.3 Uncertainty analysis

The impact velocity and initial droplet size measurements were derived from analyzing the pixel variance of the droplets in the high-speed videos using the software "Tracker". We estimate an error of ±2 pixels for the droplet size, which causes an uncertainty in the droplet diameter of less than 1%, and ±2 pixel/s for the impact velocity. As a result, the uncertainties of the impact velocities are estimated to be 0.5% to 2.0%.

The nominal resolution of the temperature readout of the IR camera sensor is 25 mK. The IR-measured temperature was calibrated for all the samples using a thin film RTD sensor (Omega, Pt100) with an accuracy of ±0.1 °C. The readouts were related to the temperature values measured by the RTD using quadratic curve fittings. Combining these errors together, the uncertainty of the IR temperature measurement is ±0.3 °C. Each experiment was repeated at least three times (five times for IR) to ensure good reproducibility of the results. The substrate was allowed to reach its steady-state temperature between the successive impact events for single-droplet impact experiments.

# 3 Results and Discussions
## 3.1 Droplet lifetime and Leidenfrost temperature

The droplet evaporation lifetimes at different initial substrate temperatures, $T_S$, for the flat and post substrates are given in Fig. 2. In general, as expected [5], the droplet lifetime decreases with an increasing initial substrate temperature before onset of film boiling. For substrate temperatures lower than 105 °C, the post substrate shows 10% to 20% shorter droplet lifetimes. In this temperature range, heat transfer is primarily conduction-limited and droplet lifetimes are dictated by droplet evaporation near the triple-phase contact line [7]. Due to the small receding contact angle on the bare aluminum surfaces, which prevents the droplet from retracting, evaporation occurs primarily after the droplet has pinned [41]. The difference in droplet lifetime on the two substrates is then mainly caused by a larger liquid-solid interface area and a lower thermal resistance of the aluminum post (as compared to bulk liquid water), which leads to higher droplet temperatures and faster evaporation [30]. For temperatures between 110 °C and 150 °C, as the droplet enters the nucleate boiling regime, the heat transfer from vaporization at the liquid-solid interface becomes increasingly important. As the substrate temperature increases, the difference in evaporation times for the different surfaces becomes vanishing small. Droplet dynamics and the cooling capacity in the nucleate boiling regime will be discussed in more detail in Section 3.3. Here, we can identify that the substrate temperatures that provide the minimum droplet lifetime for both types of substrates are around 150 °C, which correspond to the critical heat flux (CHF) temperature [42].

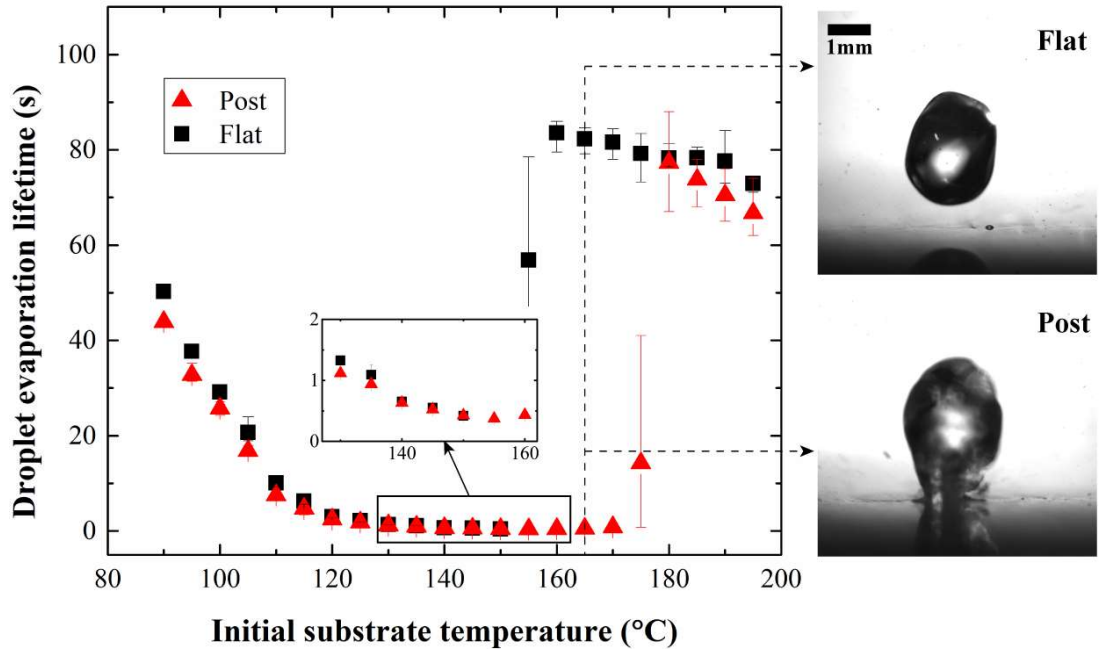

Fig. 2 Droplet evaporation lifetime on the post and flat substrates at different initial substrate temperatures for $We \approx 1.5$. The snapshots compare typical droplet shapes at $T_S = 165$ °C.

The sudden increase in evaporation time at higher temperatures indicates the transition to film boiling, with a Leidenfrost temperature $T_L \approx 160$ °C for the flat substrate and $T_L \approx 180$ °C for the post substrate, respectively. Since the data for Fig. 2 was obtained for $We \approx 1.5$, this $T_L$ is representative of the static Leidenfrost temperature. The higher $T_L$ for the post substrate is attributed to a special boiling behavior, which can be understood by looking at typical droplet shapes at $T_S = 165$ °C, as shown in Fig. 2. At this temperature, the droplet on the flat surface has entered the film boiling regime, in which a continuous vapor layer separates the liquid droplet from the hot substrate. This vapor layer acts as a thermal insulator, enabling a droplet lifetime of approximately 80 s (compared to less than 0.5 s at 150 °C). On the post substrate, however, the droplet remains attached to the post structure and remains in partial contact with the horizontal base surface. Atomization is observed around the contact line region, indicating strong vaporization at the base. The droplet shows a "candle" shape without lifting off the post and completely vaporizes in around 0.6 s. A comparison of the detailed droplet dynamics for both surfaces, including lifetime data at We ≈ 20, is provided in Supplementary Material Section 1.

This interesting boiling behavior on the post substrate is attributed to a mixed boiling mode of the droplet. On the base substrate surrounding the post, the temperature is higher than the Leidenfrost temperature of the flat surface (160 °C). The liquid-solid interface there is consequently in the film boiling mode, in which the droplet detaches from the surface. However, since the post is entirely engulfed by the droplet, its temperature is decreased by the surrounding water body, allowing for nucleate boiling rather than film boiling towards its top. The temperature decrease along the post can be estimated using an analytical model based on the heat conduction equation.

Figure 3 (a) shows the schematic of the boundary conditions. Several assumptions can be made to simplify the problem. First, we assume a 1-dimensional (1D) temperature distribution along the vertical direction due to the aluminum's high thermal conductivity and the axisymmetric design.

Second, the sidewall and top surfaces experience (nucleate) boiling and have the same constant heat convection boundary condition, which can be estimated using the lifetime, total energy (sensible and latent heat), and contact area of the droplet. Third, an isothermal boundary condition with the temperature equal to the initial steady-state substrate temperature is assumed at the bottom of the post. Additionally, we assume temperature-independence of thermophysical properties.

The heat transfer coefficient is estimated by the following steps. First, we quantify the total energy needed for vaporizing a droplet initially at room temperature $T_0$, which consists of sensible and latent heat:

$$Q_{total} = mc_p(T_{sat} - T_0) + mh_v, \qquad (1)$$

where $m$ is the mass of the droplet, $c_p$ is the specific heat, $T_{sat}$ is the saturation temperature, and $h_v$ is the latent heat of vaporization. Then, we can calculate the average heat transfer coefficient

$$h_b = Q_{total}/(\tau_{life} A_{sl}(T_s - T_{sat})), \qquad (2)$$

where $\tau_{life}$ is the measured droplet lifetime, $A_{sl}$ is the average liquid-solid contact area, and $T_s$ is the initial substrate temperature. For a post substrate at 165 °C, the assumed contact area is the surface area of the post structure (due to a "candle" shape droplet). For all the other surface temperatures, an average contact area $A_{sl}$ can be estimated from the high-speed videos. With a droplet lifetime of 0.54 s at this temperature, the estimated heat transfer coefficient is $8\times 10^4$ W/(m²K).

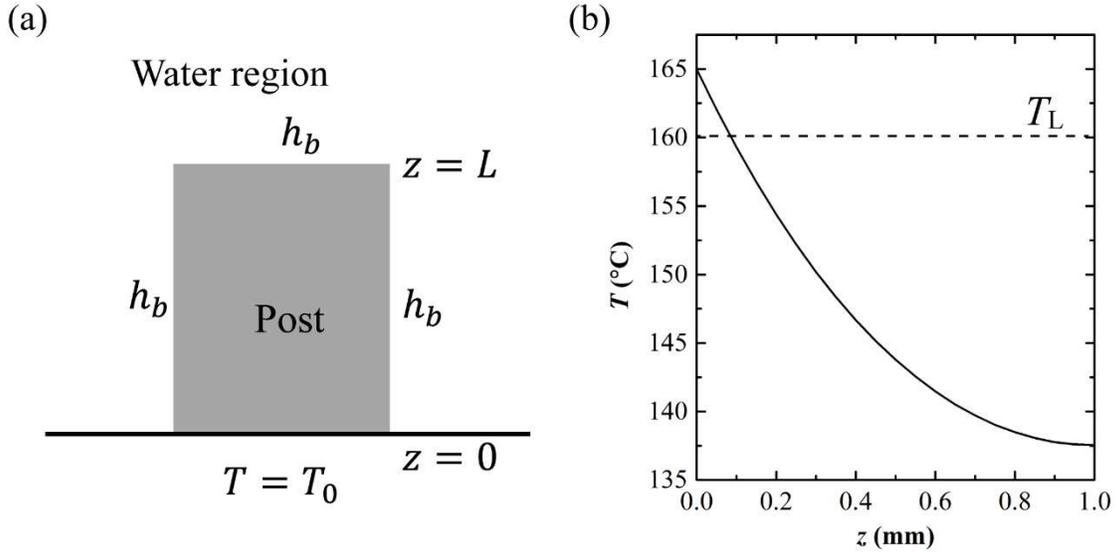

Fig. 3 Analytical model to estimate the temperature distribution along the cylindrical post. (a) The boundary conditions of the 1D heat equation. (b) Estimated temperature distribution along the post for $T_S$ = 165 °C using eq. (6). The dashed horizontal line represents the Leidenfrost temperature of the flat surface.

With these assumptions, the 1D heat conduction equation along the post can be written as,

$$k\frac{\partial^2 T}{\partial z^2} - h_b(2\pi R)(T - T_{sat}) = \frac{1}{\alpha}\frac{\partial T}{\partial t}, \qquad (3)$$

where $k$ is the thermal conductivity of the aluminum post, $R$ is the radius of the post, and $\alpha$ is the thermal diffusivity of the aluminum post. The second term on the left-hand side represents the heat convection from the sidewalls of the cylindrical post, which is determined solely by the local wall superheat ($T - T_{sat}$), since $h_b$ is assumed to be constant.

From a previous study [43], we know that the time scale of this transient process can be estimated as

$$\tau^* \approx \frac{1}{\alpha(m^2)} \approx 0.01 s, \tag{4}$$

where $m = \sqrt{\frac{2h_b}{kR}}$ is the fin parameter. This characteristic time scale of conduction is much smaller than the droplet lifetime (> 0.3 s), which means the temperature of the post becomes stable at very early times of droplet evaporation. Therefore, the heat conduction along the post can be simplified to a steady-state problem, and eq. (3) can be written as

$$\frac{d^2\theta}{dz^2} - m^2\theta = 0, \tag{5}$$

where $\theta = \frac{T - T_{sat}}{T_S - T_{sat}}$ is the non-dimensional temperature. The general solution of eq. (5) is

$$\theta = ae^{mz} + be^{-mz}, \tag{6}$$

where $a$ and $b$ are constants. The boundary condition at the bottom surface is isothermal, i.e. $\theta = 1$ at $z = 0$. The convection boundary condition on the top surface can be converted to a one-dimensional form, $k\frac{d\theta}{dz} = C_A h_b \theta$ at $z = L$, where $C_A$ is the area constant that has a value of $\pi R^2$ but a unit of 1. With these boundary conditions, the constants become:

$$a = \frac{kme^{-mL} + C_A h_b e^{-mL}}{2km \cosh(mL) - 2C_A h_b \sinh(mL)}, \text{ and} \tag{7}$$

$$b = \frac{kme^{mL} - C_A h_b e^{mL}}{2km \cosh(mL) - 2C_A h_b \sinh(mL)}. \tag{8}$$

Equation (6) along with eqs. (7) and (8) can then be used to calculate the temperature distribution along the post. Figure 3 (b) shows the temperature profile of the post for $T_S$ = 165 °C and $h_b \approx 8 \times 10^4$ W/(m²K). This bottom substrate temperature is higher than the (flat surface) Leidenfrost temperature, while the temperature drop along the post can be over 20 °C, leading to mixed boiling modes along the post height: film boiling at its base and nucleate boiling at the top. The large nucleate boiling area creates plenty liquid-solid interfaces around the post structure, which, due to the hydrophilic nature of aluminum, provide a pinning force that prevents the droplet from fully detaching from the post. As a result, the droplet body experiences competing effects from gravity, the lifting force from the bottom surface due to rapid vaporization, and the adhesion to the post surface, which eventually lead to the observed candle-like shape. The droplet always remains attached to the substrate and thus possesses a much shorter lifetime compared to the droplet on the flat surface. Similar conclusions can be drawn even when assuming non-uniform boundary conditions and/or different heat transfer coefficients at the post sides and top (see Supplementary Material Section 2).

## 3.2 Boiling regimes and droplet breakup during impact

Compared to droplet "impact" at vanishing Weber numbers, which was discussed in the previous section, droplets impacting at higher velocities show various hydrodynamic behaviors that depend on the Weber number and substrate temperature [5]. Early-stage dynamics, such as breakup, deposition, or rebound, dictate the droplet morphology and droplet-substrate contact area at longer times, which consequently has a great influence on the cooling capacity of the droplet, as will be shown in more detail in Sections 3.3 and 3.4 [44]. Fig. 4 presents time-series images for six different droplet impact and boiling modes on the post surface. Deposition was observed for droplet impact with substrate temperatures lower than 130 °C, in which the droplet surface remained smooth and no bubble or atomization occurred as the droplet spread to its maximum diameter (Fig. 4 (a), $t =$ 5.0 ms). In the deposition mode, no breakup occurred for the entire Weber number range studied in this work. However, as indicated in some previous studies, one can expect droplet breakup in the deposition mode once $We > 160$ [12]. At higher substrate temperatures, atomization (*i.e.*, the emergence of small satellite droplets) can be observed as the droplet spreads to its maximum diameter, shown in Fig. 4 (b), (c), and (d). There are two different sources for these small droplets. At relatively lower substrate temperatures ($T_S < 200°C$), the vapor layer cannot fully levitate the spreading liquid film, which causes the generation of bubbles near the contact line region. These bubbles quickly burst and lead to atomization around the rim during spreading, as shown in (b), $t = 5.0$ ms. This low-temperature atomization is known to appear more easily on hydrophilic surfaces, such as the uncoated aluminum used in this work [45]. When substrate temperatures are higher (200 °C $< T_S <$ 300 °C), the small droplets are mainly ejected from the central region (c and d), indicating that only the central area of the droplet contacts the solid surface [46, 47]. The atomization is eliminated when the substrate temperature is higher than 310 °C and the droplet enters the rebound (e) or rebound with breakup (f) modes. The substrate temperature and Weber number at the transition to these regimes mark the dynamic Leidenfrost temperature. The droplet immediately enters the film boiling regime and the vapor layer prevents the droplet from contacting the solid surface. In addition to temperature, the Weber number plays an important role in determining the breakup behavior of the droplet. For the same substrate temperature, the droplet breaks up when the impact Weber number is larger than a certain threshold. Comparing (e) and (f), we can conclude that a higher Weber number provides a larger spreading diameter, preventing the droplet from recoiling back to a single body.

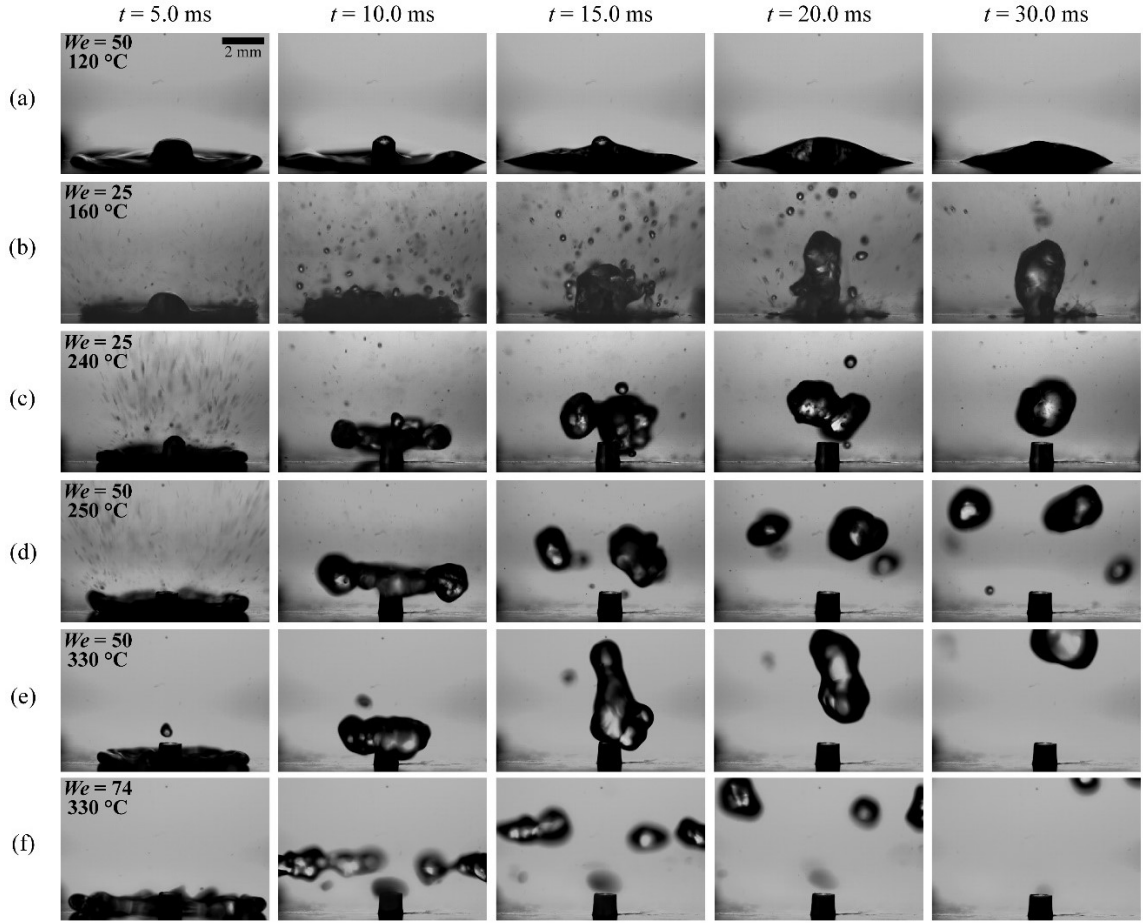

Fig. 4 Typical water droplet dynamics and boiling modes on the post substrate. (a) Deposition at $We = 50$, $T_S = 120$ °C; (b) Atomization at $We = 25$, $T_S = 160$ °C; (c) Rebound with atomization at $We = 25$, $T_S = 240$ °C; (d) Breakup with atomization at $We = 50$, $T_S = 250$ °C; (e) Rebound at $We = 50$, $T_S = 330$ °C; (f) Breakup at $We = 74$, $T_S = 330$ °C.

As seen above, the droplet impact and boiling behaviors strongly depend on Weber numbers and surface temperatures, which can be classified into five regions on a $We$-$T_S$ regime map, as shown in Fig. 5. The regime map includes representations of all six modes (where we combine deposition with and without atomization into a single region for clarity), with the detailed experimental data points provided in Supplementary Material Section 3. The solid and dashed lines represent the transitions between these regimes on post and flat substrates, respectively. Compared to the flat substrate, the regime map of the post substrate shows three major differences. First, rebound occurs at flat substrate between 160 °C to 190 °C (at certain $We$), but not for the post substrate. This difference is attributed to the additional nucleate boiling sites around the post structure that prevent the lift-off, as explained in the previous section. Second, the rebound and breakup regimes occur at slightly lower substrate temperatures and $We$ for the post substrate. This atomization-to-no atomization transition represents the onset of the dynamic $T_L$, as indicated by the blue lines in Fig. 5. The difference in dynamic $T_L$ can be explained by an increased total surface area (by the post) that decreases the dynamic $T_L$ on the structured surface [18]. Third, the droplet breakup (black dashed lines) occurs at a lower Weber number on the post substrate, especially at the temperature ranges of 170 °C to 200 °C and 290 °C to 330 °C. This observation indicates that the post structure enhances droplet breakup at multiple temperature ranges. On the post substrate, interestingly, droplet dynamics switch forth and back between breakup and rebound regions at Weber numbers ranging from 50 to 60 and at temperatures between 190 °C and 230 °C. This phenomenon is much weaker on the flat surface. This counterintuitive influence of the post structure on droplet breakup requires a detailed analysis of the droplet dynamics. Around 210 °C, rebound is sustained to higher Weber numbers than expected from dynamics on the post surface.

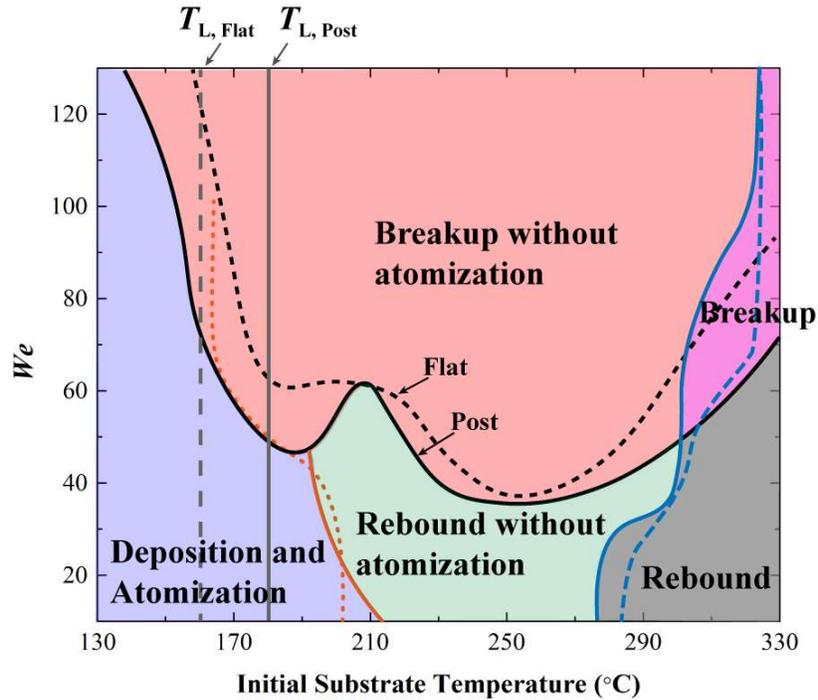

Fig. 5 $We$-$T_S$ regime map(s) for droplet impact on the post and flat substrates. Five regions are classified with colors: Deposition and Atomization (blue), Rebound without atomization (green), Breakup without atomization (red), Rebound (gray), and Breakup (pink). The solid and dashed lines represent the transitions between these regimes on post and flat substrates, respectively. The two vertical lines indicate the respective static Leidenfrost temperatures (for $We \rightarrow 0$). The blue curves indicate the dynamic Leidenfrost temperatures.

Figure 6 provides insights into the breakup-rebound-breakup transition on the post substrate. Shown is the temporal evolution of droplet dynamics in side and top views, as well as schematic representations of the processes. The three droplets impact the substrate with the same $We$ number, but at different $T_S$. At the lower temperature (≈190 °C), the center of the droplet, which touches the post, experiences nucleate boiling, pinning it in place due to its high wettability, as discussed in Section 3.1, while the spreading rim is supported by the vapor layer (Fig. 6 (a), $t$ = 5.6 ms). Consequently, the rim experiences very little resistance during spreading, leading to the development of a thin liquid film with decreasing thickness, which eventually breaks up. Nucleate boiling and droplet pinning on the post surface, along with the low-temperature breakup mode, lead to the main droplet being attached to the post and satellite droplets surrounding it. At high temperatures (> 230 °C), the post substrate supports the transition boiling regime where only the central region shows atomization [27, 28] during spreading (Fig. 6 (c), $t$ = 4.0 ms). However, the central region quickly turns to film boiling before 5.4 ms, as the surface temperature recovers due to the continuous heat supply from the substrate. The large droplet detaches from the post during the spreading process. Outward inertial forces and the generated vapor work together to drive the liquid film away from the post and create an expanding central cavity, as seen in (c) at $t = 5.4$ ms. Then, as the liquid film disconnects, the whole droplet breaks up into several small droplets, leaving the center post dry. In between these two breakup modes, a transition mode exists (200 < $T_S$ < 220 °C), in which the droplet does not break up and lifts off the post structure as a whole (Fig. 6 (b)). In this mode, the droplet is also in the transition boiling regime, but it takes longer for the liquid film to detach from the central post region due to a lower substrate temperature. Instead of a detachment from the post during the spreading process, the liquid now de-wets the post during the receding stage. As the rim retracts, the vapor generation at the post is too weak to provide sufficient outward momentum to counteract the inertia from the retracting liquid. Therefore, upon recoil, the droplet re-connects with the post and lifts off. Although this discontinuity of breakup can be found only in a small range of *We* numbers, it reveals that the interplay between surface morphology and temperature plays a critical role in determining droplet dynamics.

We also investigated the droplet impact and boiling mode at a smaller droplet size (1.8 mm), and found that the breakup transition Weber numbers are very similar to the original 2.7 mm-droplet. This result indicates that the droplet breakup on the post substrate is still highly determined by the relative strength between inertia and capillary forces. More information on this smaller-diameter droplet impact can be found in Supplementary Material Section 4.

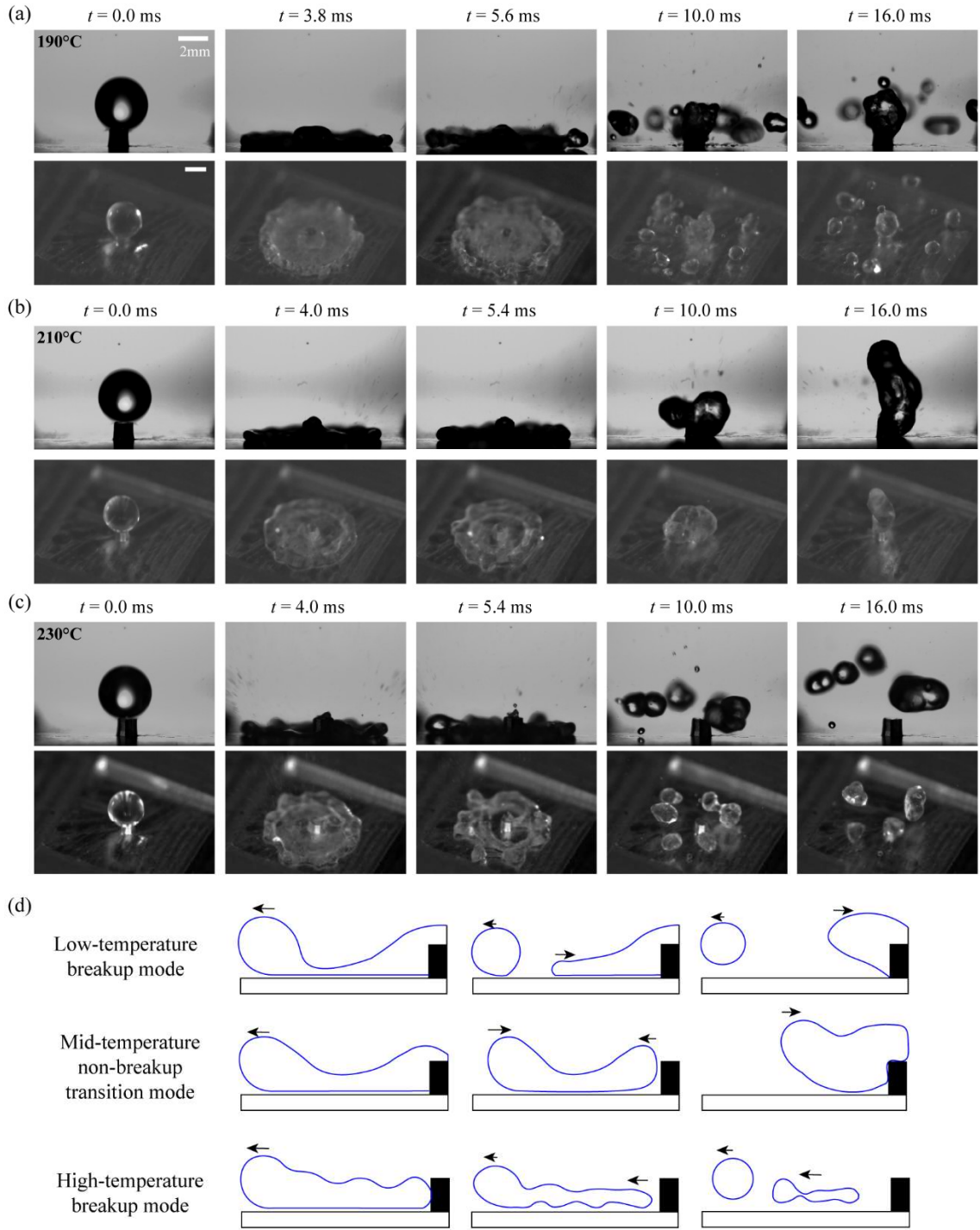

Fig. 6 Comparison of droplet (breakup) dynamics for different substrate temperatures (all at $We \approx 57$). (a) Low-temperature breakup mode at $T_S = 190$ °C. (b) Mid-temperature non-breakup transition mode at $T_S = 210$ °C. (c) High-temperature breakup mode at $T_S = 230$ °C. (d) Schematics of the different breakup mechanisms. The top view snapshots are recorded at an angle of approximately 20° from the vertical.

After investigating the two different breakup mechanisms on the post surface, we can now revisit the $We$-$T_S$ regime map (Fig. 5) and better understand the transitions in droplet modes for the flat and post surfaces, respectively. The post substrate shows an earlier breakup (smaller *We*) between 170 °C and 200 °C, which corresponds to the low-temperature breakup mode. For the temperature range of 290 °C to 330 °C, the high-temperature breakup mode also causes breakup at a lower *We*. Supplementary Material Section 5 visually compares droplet evolutions on the post and flat substrate at 180 °C and 320 °C, respectively, and provides more explanations on the (non-) breakup dynamics.

### 3.3 Cooling capacity in the nucleate boiling regime

After discussing the droplet hydrodynamics, we now turn our focus to the more practical aspect of this study: quantifying the influence of the post on the droplet's ability to efficiently cool the surface. Measuring the transient surface temperature of impacting droplets is a common approach for quantifying the cooling capacity of different surfaces [48, 49]. In this study, we recorded the backside (bottom) surface temperature of the two substrates using high-speed IR imaging to compare their respective cooling efficiencies. The bottom surface temperature, although it experiences a lag in response time, provides a better metric to quantify the cooling capability due to its better resemblance to real-world droplet impingement applications (such as quenching) and has been used by a number of previous studies [50-52]. As shown in Fig. 2, the droplet lifetime on the two substrates becomes very similar for 120 °C < $T_S$ < 150 °C. While droplet lifetime measurements provide temporal information on phase change dynamics, IR thermography provides both temporal and spatial information on the cooling performance. The measurement of the bottom surface temperature can thus provide better and direct evidence to compare the cooling capabilities between the two substrates in the nucleate boiling regime.

Fig. 7 shows the area-weighted average temperature evolution of the impact area (8 mm × 8mm) at the bottom of the substrates for four different initial temperatures. We choose the non-dimensional time $t^* = t/t_0$, where $t_0$ is the time of full evaporation on the post substrate (compare to Fig. 2), to present the temperature history. Dimensional temperature data can be found in Supplementary Material Section 6. For all four cases, the temperature decreases rapidly as the droplet impacts the surface and gradually recovers due to a continuous heat supply from the two heating blocks and the reduced heat loss to the vanishing droplet. A larger temperature drop indicates that the droplet removes more heat from the substrate and hence has a higher cooling capacity. The post surface has a 40% and 22% higher maximum temperature drop for $T_S$ = 110 °C and $T_S$ = 130 °C, respectively. However, as the initial substrate temperature increases, the difference in cooling capacity becomes negligible between the post and flat surfaces (< 8% difference in temperature drop). These results match those of the droplet lifetime shown in Fig. 2, where droplets evaporate faster on the post than the flat substrate at 110 °C and 130 °C, but have a similar lifetime at 150 °C. The two substrates show a tremendously different temperature history at 165 °C, though, which is attributed to the Leidenfrost effect on the flat substrate where the vapor layer significantly blocks the heat transfer. As discussed in Section 3.1, the post enables droplet pinning at this temperature, leading to significantly enhanced cooling.

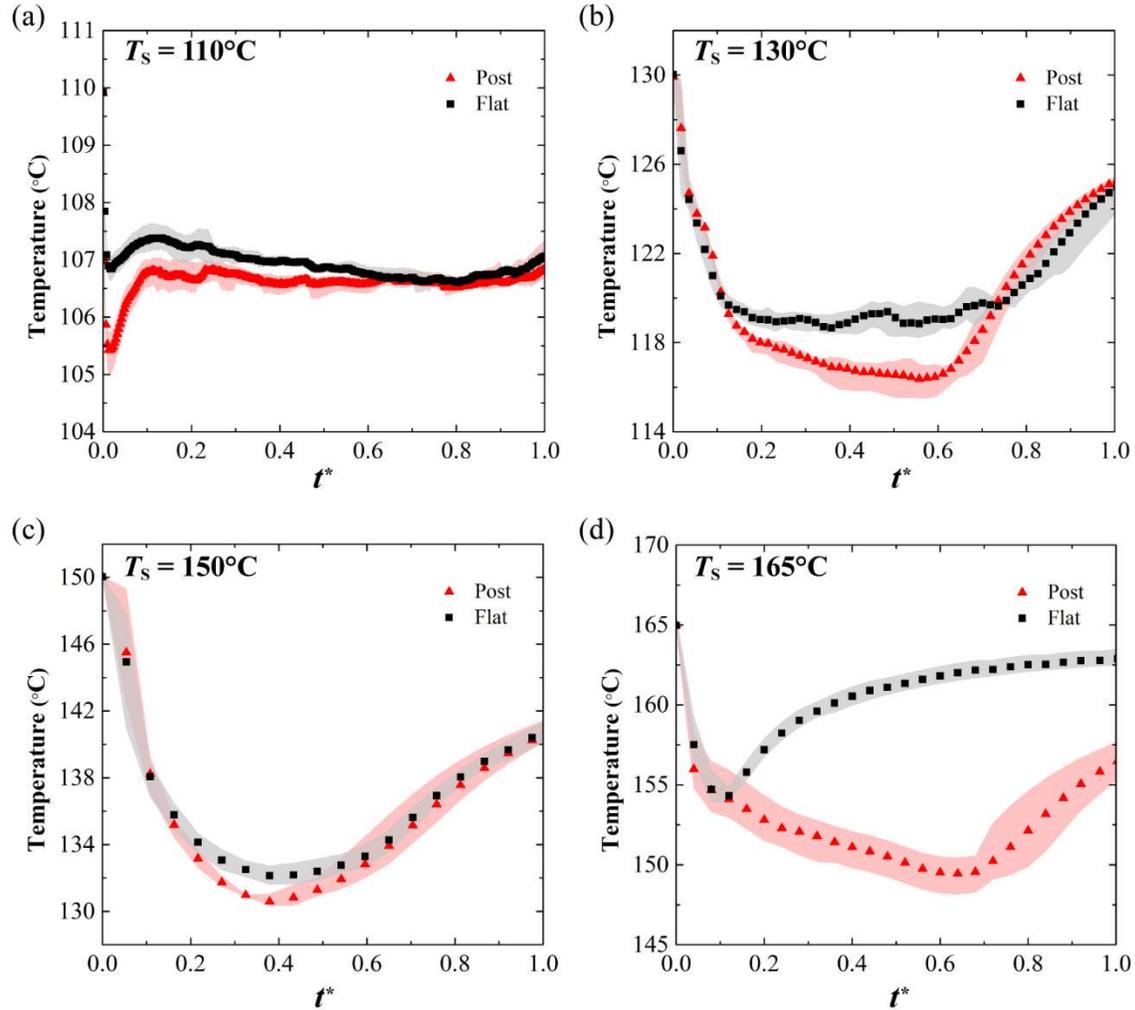

Fig. 7 Area-weighted average bottom-surface temperatures for (a) $T_S = 110$ °C, (b) $T_S = 130$ °C, (c) $T_S = 150$ °C, and (d) $T_S = 165$ °C within the droplet lifetime on the post substrate. The x-axis is the non-dimensional time $t^* = t/t_0$, where $t$ is the real time and $t_0$ is the droplet lifetime on the post substrate at the respective initial temperature (compare to Fig. 2). For all the experiments, droplets were initially at room temperature ($\approx 25$ °C) and impacted the substrates with $We \approx 20$. Schaded areas represent the error interval from five measurements.

To investigate the differences of the temperature curves in the nucleate boiling regime, and especially the nearly identical profiles in Fig. 7 (c), we analyze the bubble generation and the droplet profile through high-speed imaging. Detailed information is given in Supplementary Material Section 7. Here, we summarize the important findings. At $T_S = 110$ °C, the droplet on the flat substrate experiences gentle nucleate boiling without significant lateral expansion. On the contrary, the droplet on the post substrate undergoes foaming that leads to an expansion of the droplet-solid contact line. At $T_S = 150$ °C, both substrates show similar boiling behavior where the bubble generation is vigorous and chaotic. Due to the distortion of the droplet, the post structure is partially exposed, reducing the influence of the (nominally) additional liquid-solid contact area compared to the flat substrate. This observation explains the minimal difference in the substrate temperatures shown in Fig. 7 (c). In conclusion, the bubble generation and boiling behavior on the two substrates become increasingly alike between $T_S$ = 110 °C and 150 °C, which eventually leads to similar droplet lifetimes and temperature drops at $T_S$ = 150 °C.

As mentioned above, the post surface modifies droplet dynamics and nominally has a higher contact, or heat transfer, area than the flat surface. On the other hand, the cylindrical post (= fin) also poses an additional thermal conduction resistance. A basic heat transfer analysis can help better determine the most influential factors and quantify the cooling capacity enhancement of the post structure.

For the flat surface, we can estimate the convection heat transfer to the droplet with

$$q_{flat} = h_b A_{sl}(T_S - T_{sat}), \qquad (9)$$

where $A_{sl}$ is the average solid-liquid interfacial area of the droplet on the flat surface and $h_b$ is the same convection coefficient that we used in Section 3.1. The heat transfer rate on a post substrate can be estimated as

$$q_{post} \approx h_b A_p (T_{top} - T_{sat}) + h_b (2\pi R L)(T_{ave} - T_{sat}) + h_b (A_{sl} - A_p)(T_S - T_{sat}), \qquad (10)$$

where $A_p$ is the top surface area of the post, $T_{top}$ and $T_{ave}$ are the temperatures at the top of the post and the average temperature of the sidewall of the post, respectively, which can both be calculated using eq. (6). Comparing eqs. (9) and (10) for substrate temperatures of 110 °C, 130 °C, and 150 °C, the post substrate has 24%, 18%, and 15% higher heat transfer rates, respectively, compared to the flat substrate. These results indicate that the improvement of the cooling capacity observed in Fig. 7 stems primarily from the additional contact area between droplet and solid. Due to the high thermal conductivity of the aluminum, the conduction resistance within the post is minimal and the addition of the post can effectively enhances the cooling capacity on the substrate.

Motivated by the post's general superiority in achieving a cooling capacity for a single-droplet impact, we further explored the temperature evolution during multi-droplet impact to identify whether the improvement can be maintained. Detailed experimental data for multiple droplets impacting the substrates are provided in Supplementary Material Section 8. Briefly, the results show that the post substrate can still sustain a higher cooling capacity even for multi-droplet impact for substrate temperature between 110 °C and 165 °C, as long as the impact frequency (or volume flow rate) is low enough to prevent flooding.

### 3.4 Droplet impact on inclined surfaces

So far, the hydrodynamics and the cooling capacity of the two surfaces were investigated on horizontal substrates with perpendicular droplet impingement. However, in real applications, droplets often impact solid walls at various inclination angles [53]. Previous studies have explored the hydrodynamics of non-isothermal droplet impact on inclined surfaces, including the influence of Leidenfrost effects [53, 54], wettability effects [55], and surface roughness effects [56]. On a smooth surface, the inclination can cause the droplet to slide or bounce off the substrate, decreasing the local heat transfer at the initial impact location [54]. For a post substrate, which already positively influences impact and heat transfer characteristics when placed horizontally, one can expect even stronger benefits when mounted at an angle.

We investigated droplet impact on the two substrates inclined at 30° from the horizontal and at different initial temperatures. Fig. 8 provides snapshots of droplet dynamics and the back-side temperature distribution for droplet impact at an initial temperature of $T_S$ = 145 °C. As discussed above, there is a slight delay in the IR signature due to the time associated with heat diffusion through the substrate. Nonetheless, the temperature profile still provides an adequate representation of the local cooling as the droplet moves across the top of the surface. On the flat substrate, as expected, the droplet slides down the incline during the spreading and recoil process and finally falls off the substrate. The temperature profile shows an elongated cold spot and that keeps shifting to the left (downhill). On the post substrate, the droplet first spreads asymmetrically due to gravity with the uphill portion of the droplet pinned to the post structure ($t$ = 20.0 ms). Then, driven by capillary forces and the pinning provided by the post, the droplet returns to the impact location and symmetrically covers the post structure ($t$ = 44.0 ms). The temperature profile shows a low-temperature region that evolves radially with time, indicating effective cooling around the impact area.

Additional results for $T_S$ = 160 °C are provided in Supplementary Material Section 9. Briefly, the droplet still pins at the post structure and avoids detachment from the substrate. But instead of a symmetric droplet shape, the droplet attaches to the left side of the post structure. On the flat substrate, the droplet shows a downward sliding during the spreading and recoil and then directly bounces up from the surface due to Leidenfrost dynamics. We define the Leidenfrost temperature on the inclined surface to be the substrate temperature at which droplet starts to slide off at We ≈ 1.5. The $T_L$ for the flat substrate and post substrates at a 30° inclination are approximately 145 °C and 165 °C, respectively.

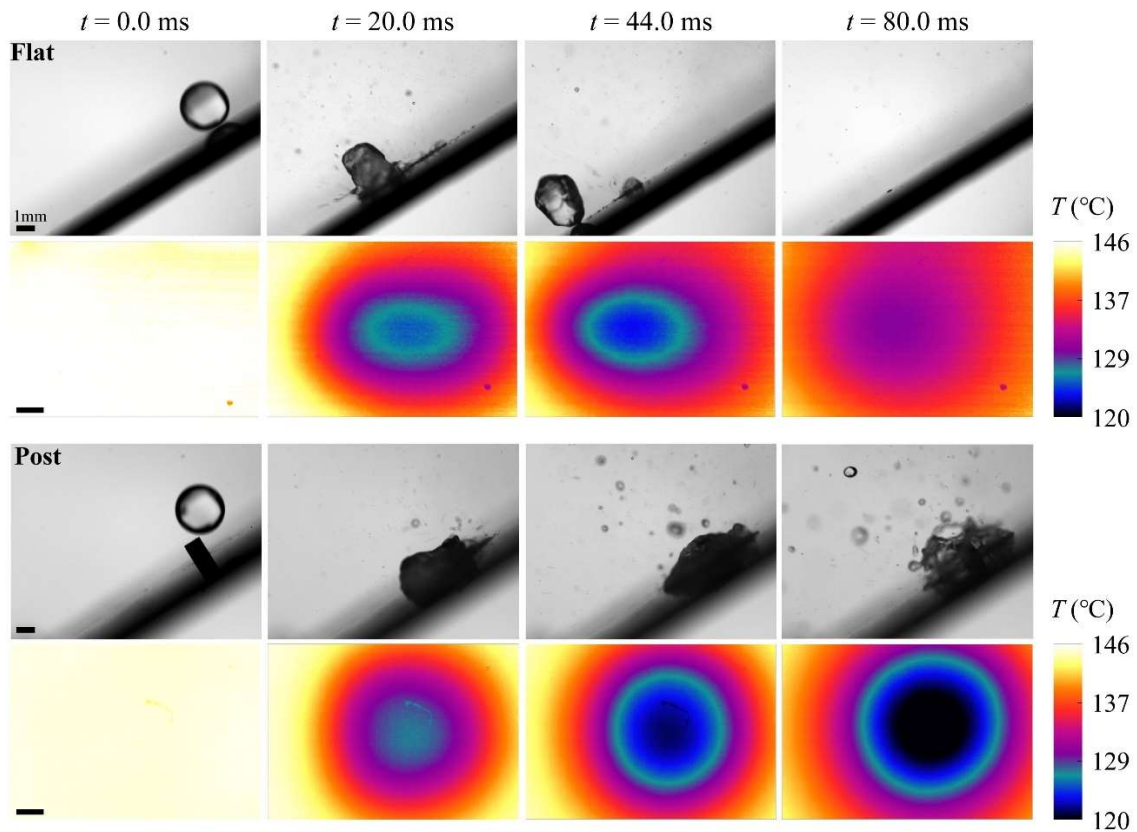

Fig. 8 High-speed and IR imaging sequence of droplets impacting on a flat (top) and a post (bottom) substrate inclined at 30° at an initial temperature of $T_S$ = 145°C and We = 65.

The area-weighted average temperature history of the impact area (8 mm × 8mm) is presented in Fig. 9. At $T_S$ = 145 °C (nucleate boiling regime), the post increases the maximum temperature drop by 51%. However, at a similar substrate temperature on a horizontal surface (Fig. 7(c)), the improvement from the post is minimal. At 160 °C (film boiling regime), the enhancement in cooling capacity is even stronger: the post increases the maximum temperature drop on the inclined surface by 180%. The mixed boiling mode and pinning on the post prevent the droplet from sliding/bouncing off the substrate. This result shows that for both the nucleate boiling and transition boiling regimes, the post structure can significantly increase the local heat transfer of the droplet impacting an inclined surface.

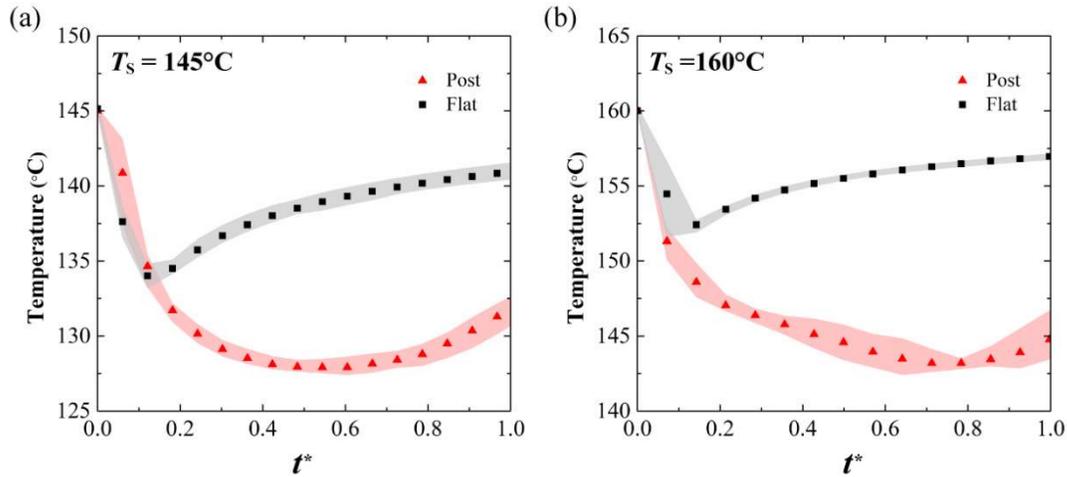

Fig. 9 Temporal evolution of the area-weighted average bottom surface temperature for droplet impacting the two surfaces at (a) $T_S$ = 145°C and (b) $T_S$ = 160°C and $We$ = 65 at an inclination angle of 30° from the horizontal.

So far, we have discussed droplet dynamics and cooling capacities for droplets centrally impacting the post right. In real-world droplet impingement applications, however, off-center impact is more likely, for which the benefit of the post structure, which prevents sliding and bounding of the droplet, might be diminished. To provide guidelines on the applicability of millimetric surface patterning, we also investigated the off-center impact on the post substrate, and found that the post structure (1 mm in diameter) can still prevent the sliding of the droplet with an impact location of up to 1.6 mm (horizontal distance) downhill from the center of the post. Detailed information on droplet dynamics for off-center impact is provided in Supplementary Material Section 9.

In addition to the here-presented inclination angle of 30° and Weber number of $We$ = 65, we also investigated the effect of droplet pinning on the cooling capacity for different inclination angles, substrate temperatures, and Weber numbers, shown in Fig. 10. We found that the droplet pinning from the post (red triangles) provides enhanced cooling for a smaller range of inclination angles as the temperature increases. At an inclination angle of 40°, due to the large droplet deformation induced by gravity, the pinning only occurs at low temperatures and small $We$. However, this still enhances the cooling considerably, since droplets directly slide off a flat surface at 135 °C and tilted by 40°.

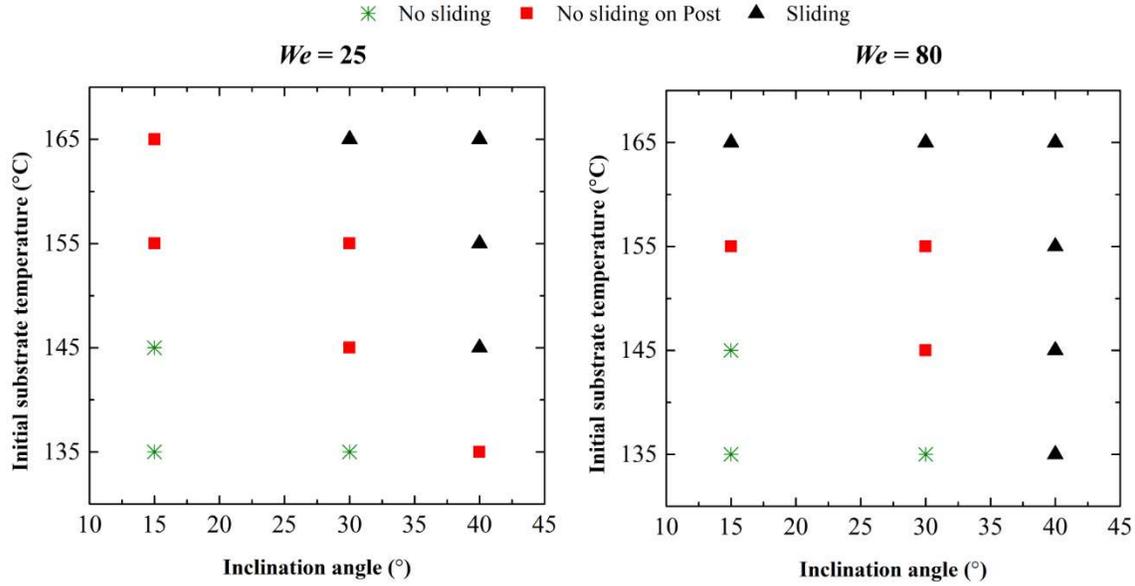

Fig. 10 Droplet sliding regime maps at different substrate temperatures and inclination angles. Green stars represent no sliding at both substrates (flat and post), Red rectangles represent no sliding on the post substrate only (while droplets slide on the flat substrate), and the black triangles represent droplet sliding on both substrates.

## 4 Conclusion

This study experimentally explored hydrodynamics and heat transfer during droplet impact on a millimetric heated post. Detailed analyses have been carried out on droplet lifetime, boiling modes, and cooling capacity. We can draw the following conclusions:

1) The post substrate shows a 20 °C higher Leidenfrost temperature compared to the flat substrate, which is attributed to a mixed boiling mode and additional pinning induced by the post.

2) Six different droplet impact and boiling modes were identified for initial substrate temperatures ranging from 120 °C to 330 °C and Weber numbers between 10 and 120: 1) deposition, 2) atomization, 3) rebound with atomization, 4) breakup with atomization, 5) rebound, and 6) breakup. $We$-$T_S$ regime maps were constructed to quantify the differences in impact and boiling modes between the post and flat substrates. In general, the post enhances droplet breakup. Unexpectedly, though, a non-breakup transition region exists for the post surface within a small temperature range between the two breakup modes, which is caused by a complex interplay of inertial and pinning forces.

3) The post substrate has a higher cooling capacity than the flat substrate between 110 °C and 130 °C due to a larger liquid-solid interface area and an expanding droplet profile. However, the enhancement becomes negligible as the temperature approaches the Leidenfrost temperature of the flat surface.

4) On an inclined surface, the post substrate successfully prevents the droplet from sliding or bouncing off the substrate, which leads to a 51% to 180% increase in local temperature drop.

Overall, we show that droplet impact on a single post considerably improves the interfacial heat transfer by influencing droplet dynamics and boiling behavior. The post structure provides an enlarged contact area and a delayed Leidenfrost effect by pinning the droplet. This pinning also assists in droplet breakup and prevents the droplet from sliding off the surface for impact on an inclined substrate. These large-structured surfaces can potentially increase the cooling performance in droplet impingement cooling applications [57] and prevent droplet bouncing for inclined impingement surfaces [58].

## Declaration of Competing Interest

The authors declare that they have no known competing financial interests or personal relationships that could have appeared to influence the work reported in this paper.

## Acknowledgments

The authors would acknowledge Kiersten Horton for assisting with sample fabrication. This material is based upon work supported by the National Aeronautics and Space Administration (NASA) under Grant No. 80NSSC20K0072 issued through the Early Career Faculty (ECF) Program. The authors also acknowledge the use of instruments and staff assistance from the Spartan Light Metal Products Makerspace at Washington University in St. Louis.

# Droplet impact and Leidenfrost dynamics on a heated post: Supplementary Material

**Junhui Li[1], Patricia Weisensee[1,2]**

[1]Mechanical Engineering and Materials Science, Washington University in St. Louis, St. Louis, MO 63130

[2]The Institute of Materials Science & Engineering, Washington University in St. Louis, St. Louis, MO 63130

## 1  Delay of the Leidenfrost effect on the post substrate

Fig. S1 shows the detailed droplet dynamics on the two substrates at $T_S$ = 165°C and $We$ =1.5. The corresponding videos can be found in Supplementary Video 1. On the flat surface, after the spreading process ($t$ = 5.0 ms), the droplet slowly de-wets the surface (seen in the video) and fully lifts off around $t$ = 60.0 ms. Then the droplet enters the Leidenfrost (film boiling) regime and departs from the initial impact location. The vapor layer in the Leidenfrost regime blocks the heat transfer and extends the droplet lifetime to around 80 s. On the post substrate, the droplet covers the post after impact and the liquid film always covers the post structure ($t$ = 5.0 ms). After $t$ = 50.0 ms, the droplet shows a "candle" shape until it fully evaporates. Due to a thin liquid layer and strong local vaporization, atomization is observed around the contact line region (seen in the video). The droplet remains pinned at the post structure and vaporizes quickly with a lifetime of approximately 0.6 s.

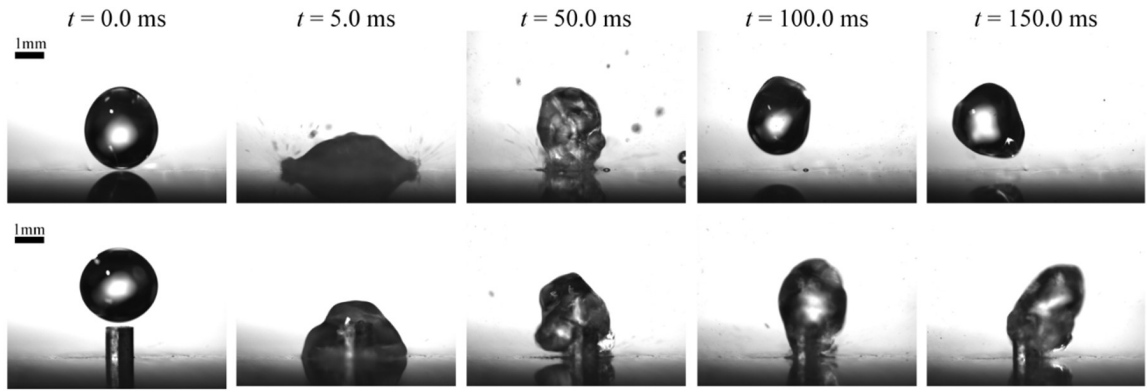

Fig. S1 Image sequence of droplet impact on a flat (top) and a post (bottom) substrate at $T_S$ = 165°C.

We also investigated the droplet lifetime at $We$ = 20, shown in Fig. S2. Except for a smaller droplet lifetime for both substrates in the low temperature range due to an increase in spreading area, the results at $We$ = 20 are very similar to the results at vanishing Weber number ($We$ = 1.5). Because the capillary time scale $\sqrt{\rho D_0^3/\sigma} \approx 15\ ms$ is much smaller than the droplet lifetime, the $We$ number should have very little influence on the droplet lifetime [1]. The Leidenfrost temperatures of the post and flat surface remain the same (180°C and 160°C).

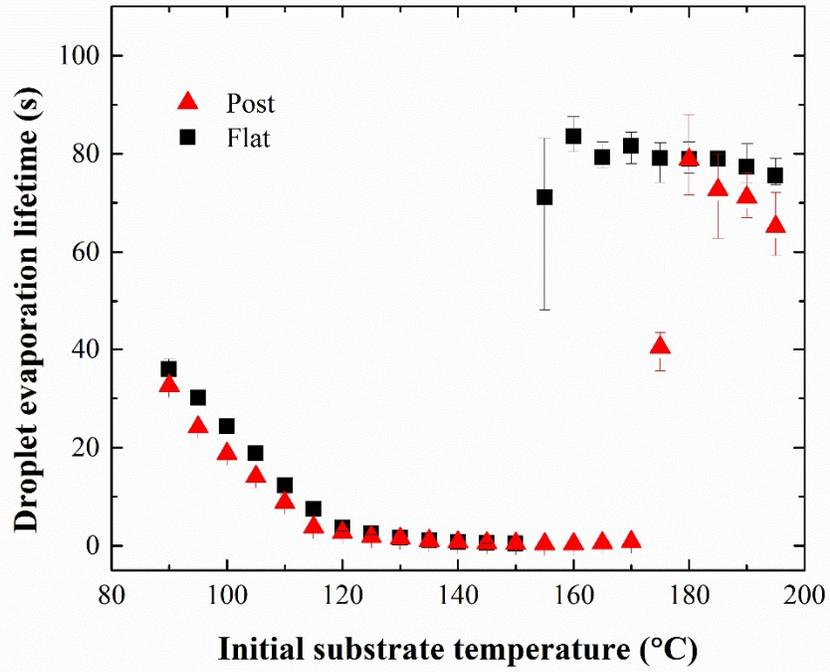

Fig. S2 Droplet evaporation lifetime on the post and flat substrates at different initial substrate temperatures for $We \approx 20$.

## 2 Analytical model with different assumptions

To show that the absolute value of the heat transfer coefficient does not influence the general conclusion of a "mixed boiling mode" along the post, we compared the temperature profile for different heat transfer coefficients. Previously, we assigned a uniform heat transfer coefficient of $8\times10^4$ W/(m²K) at the side and top. However, even with different boundary conditions at the top and side, we can still conclude that the post structure undergoes nucleate boiling. For example, we can set the heat transfer coefficient to be $2\times10^4$ W/(m²K) at the top and assign the heat transfer coefficient on the side to be $6\times10^4$ W/(m²K). These values would result in a temperature distribution that is shown in Fig. S3(a). Compared to the original results, the temperature on the top of the post increases by around 5°C. However, the general conclusion, namely that the post structure undergoes nucleate boiling, is the same as with the original assumption. We also explored the influence of using various heat transfer coefficient values (same on all surfaces), whose results are shown in Fig. S3(b). For the heat transfer coefficient ranging from $1.5\times10^4$ to $12\times10^4$ W/(m²K), the temperatures of the post are all smaller than flat surface $T_L$ and fall into the nucleate boiling regime.

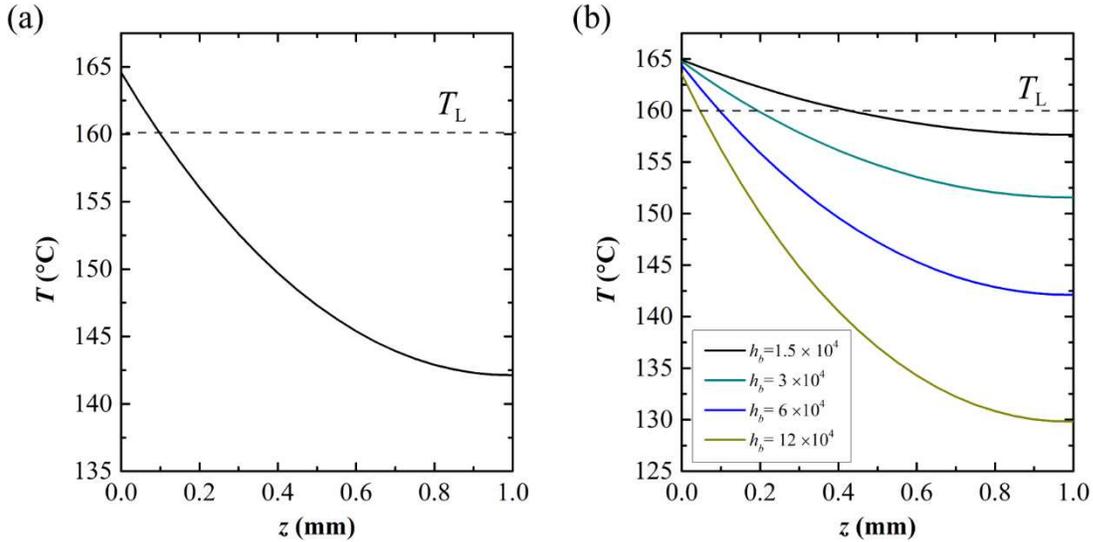

Fig. S3 Estimated temperature distribution on the post structure using (a) different boundary conditions at the top ($2\times10^4$ W/(m²K)) and side ($6\times10^4$ W/(m²K)) of the post, and (b) identical boundary conditions on all surfaces, but different heat transfer coefficient ($h_b$) values. The unit of $h_b$ is W/(m²K). The dashed lines represent the flat surface $T_L$.

# 3 Regime maps with the experimental data

The regime maps with the experimental data corresponding to Fig. 5 in the main manuscript are provided in Fig. S4. Six droplet impact and boiling modes can be found for both flat and post substrates at different $We$ and $T_S$. In addition to the transitions between the five regions that we already introduced in the main manuscript, the difference of each single mode between the two substrates based on the data points can be observed in these plots.

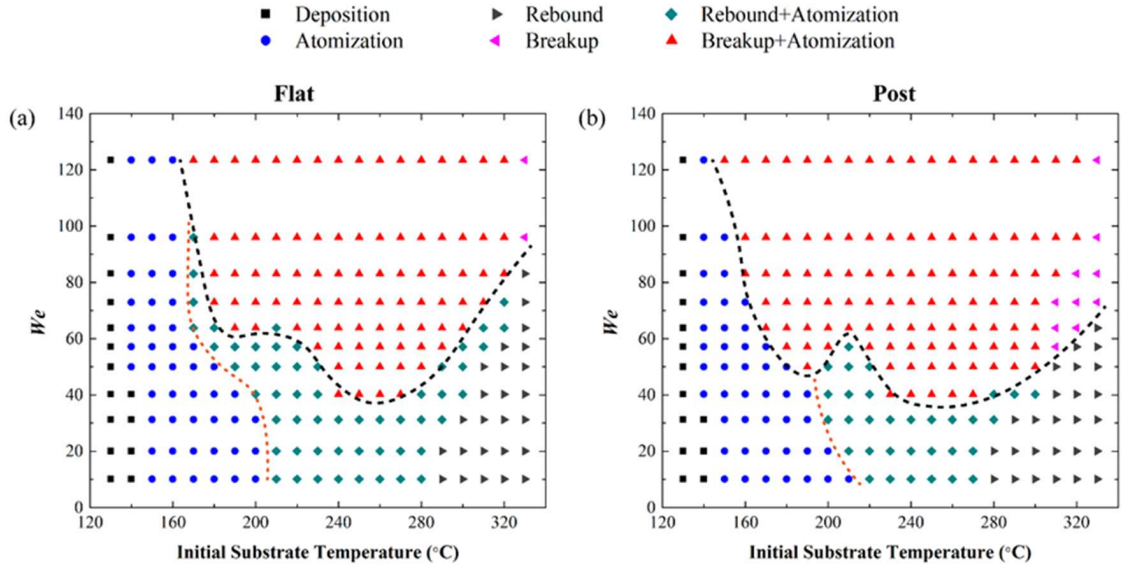

Fig. S4 $We$-$T_S$ regime maps for droplet impact on (a) a flat substrate and (b) a post substrate. Six different impact and boiling modes are observed. The orange dashed line is a visual guide for the transition between droplet rebound and non-rebound regimes. The black dashed line represents the transition between rebound and breakup regimes.

# 4 Effects of droplet size

The experimental results that we present in this study are based on a droplet size of 2.7 mm. The droplet-to-post ratio 2.7 is chosen based on several reasons. First, the droplet should easily cover the post structure and maintain an axisymmetric shape (not lean to one side), which requires a ratio larger than 1.5. Second, the post structure should still be large enough to make significant changes to the static droplet shape or dynamic behavior, which favors a ratio smaller than 3. Third, for a ratio ranging from 1.5 to 3, we would like to maximize the droplet volume to acquire the largest cooling capacity. Therefore, a droplet volume of 2.7 mm, which is easily generated by a 25-gauge needle, was chosen for the 1 mm post substrate.

In this section we present results from experiments using a smaller needle size to generate droplets at $1.8 \pm 0.05$ mm and select representative surface temperatures of 200 °C and 250 °C to investigate the droplet impact and boiling modes on a post substrate (shown in Fig. S5). At 200 °C, the smaller droplet already shows a rebound at $We = 23$. However, as seen in Fig. 5, the original 2.7 mm-droplet rebounds at $We = 31$. This difference indicates that the smaller droplet size provides an early rebound, which can be attributed to a smaller transient cooling effect that increases the vapor thickness. For droplet breakup, we found that the breakup transition Weber numbers are very similar between the 2.7 mm and 1.8 mm droplets. This result indicates that the droplet breakup on the post substrate is still highly determined by the relative strength between inertia and capillary forces.

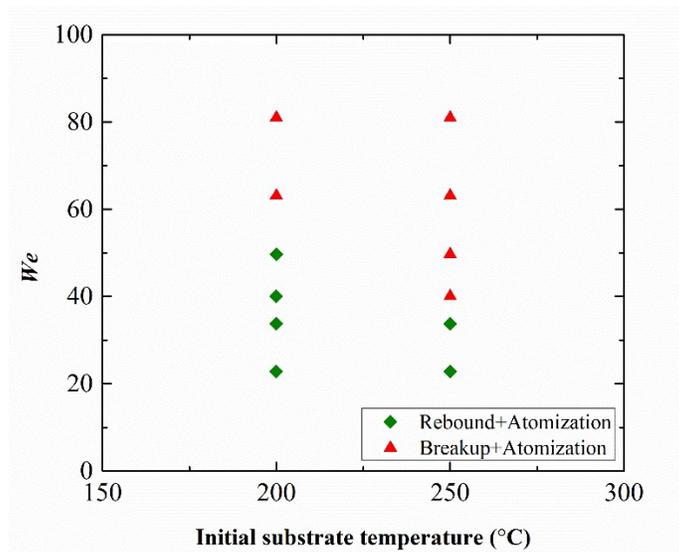

Fig. S5 Regime map of 1.8 mm droplets impacting the post substrate.

~ 5 ~

# 5 Enhancement of droplet breakup on the post substrate

Fig. S6 shows the droplet dynamics on the flat and post surfaces at 180°C and 320°C, respectively. On the flat substrate, the droplet spreads to a certain maximum diameter (Fig. S6 (a) $t$ = 6.0 ms) and then fully recoils and lifts off. On the post substrate, however, the droplet spreads to a slightly larger maximum diameter (Fig. S6 (b) $t$ = 6.0 ms) and finally ends up with a central droplet attached to the post surrounded by satellite droplets (Fig. S6 (b), $t$ = 16 ms). We hypothesize that the volume of the post at the center causes the droplet to stretch to a larger spreading diameter than on the flat surface, which leads to a thinner film that enhances the breakup. Fig. S6 (c) and (d) correspond to the high-temperature breakup mode. On the flat substrate, the droplet maintains a thin liquid layer across the entire width of the spread droplet, as shown in Fig. S6 (c) $t$ = 6.0 ms. This thin liquid layer prevents the disintegration of the liquid body, which enables non-breakup rebound. However, on the post substrate, we notice a separation of the liquid film and eventual breakup where the post "pierces" the liquid film. This separation is caused by the vapor generation at the post surface that leads to a cavity expanding radially from the center (Fig. S6 (d) $t$ = 6.0 ms).

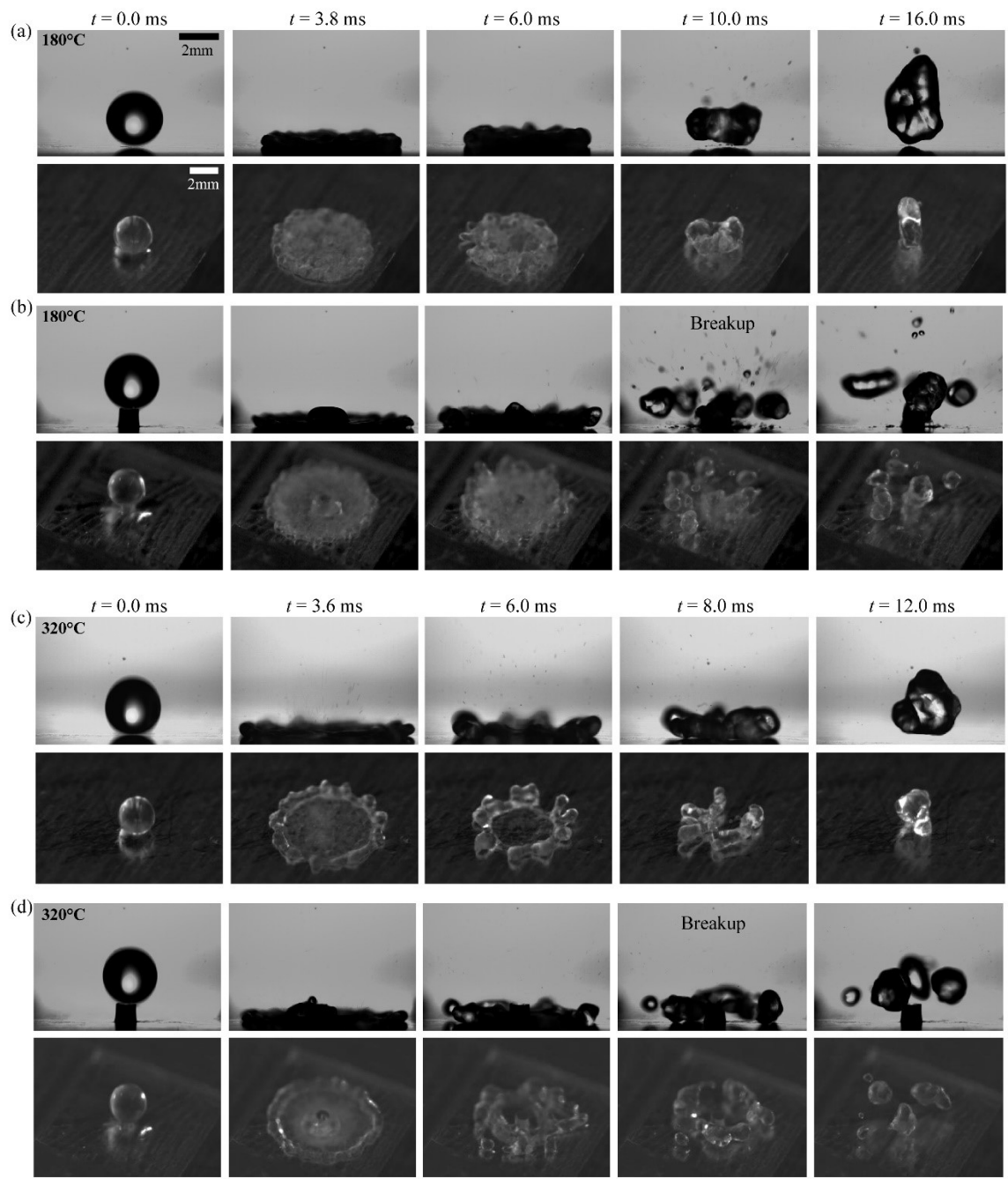

Fig. S6 Image sequence of the droplet impact at $We \approx 57$ on (a) a flat substrate at $T_S = 180°C$, (b) a post substrate at $T_S = 180°C$, (c) a flat substrate at $T_S = 320°C$, and (d) a post substrate at $T_S = 320°C$.

# 6 Cooling capacity of single droplet impact

As Fig. 5 in the main provides the temperature history within the lifetime of the droplet using a dimensionless time that is related to the droplet lifetime on the post, here we present the droplet temperature history using the real time after impact. Except for $T_S = 110$ °C, where the droplets are still at their early evaporation time, the other three cases show a strong recovery of the substrate temperature near the end due to their small droplet lifetimes.

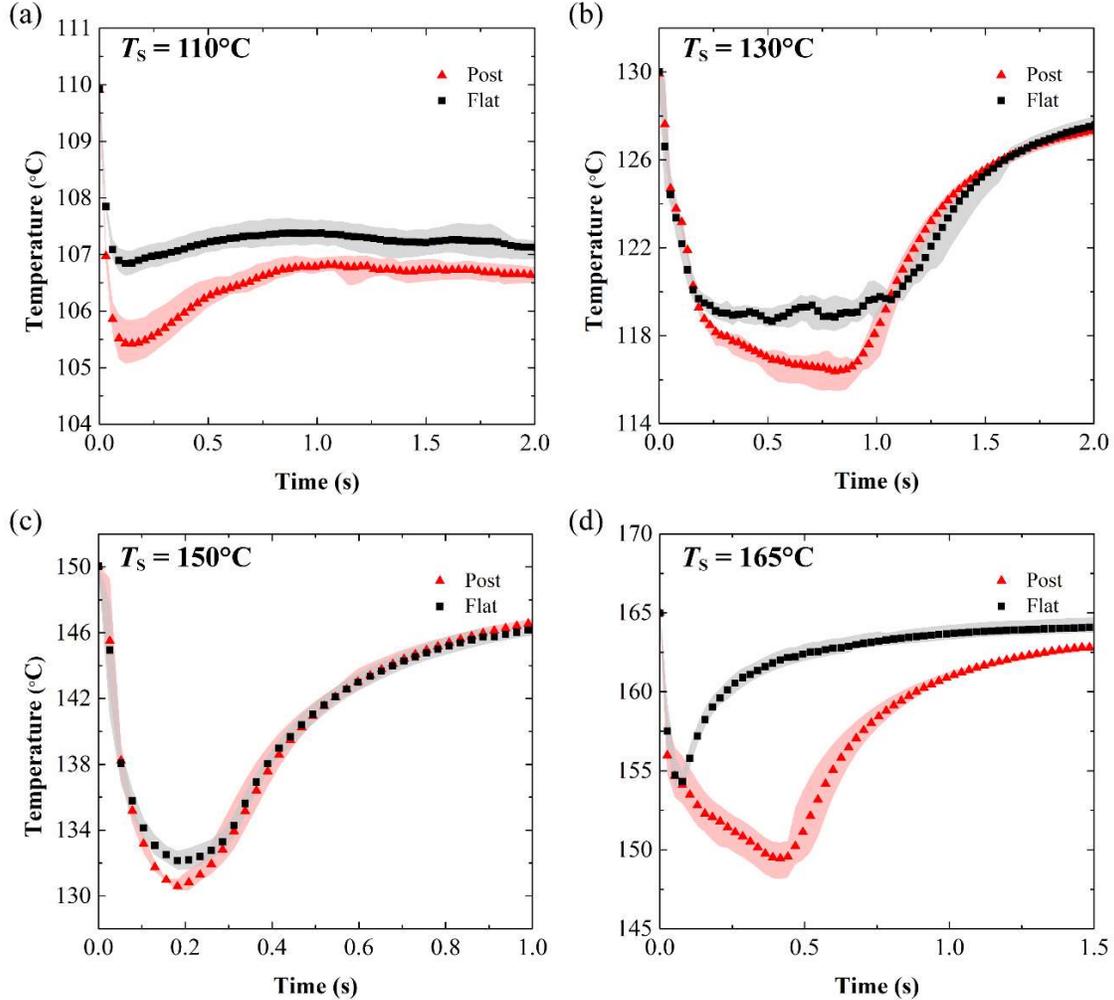

Fig.S7 Area-weighted average bottom-surface temperatures for (a) $T_S = 110$ °C, (b) $T_S = 130$ °C, (c) $T_S = 150$ °C, and (d) $T_S = 165$ °C. For all the experiments, droplets were initially at room temperature ($\approx 25$°C) and impacted the substrates with $We \approx 20$.

When considering the "optimum parameters" for cooling capacity, the addition of the post clearly indicates a strong enhancement. Additionally, the cooling capacity will be strongly influenced by the thermal conductivity of the substrate. For the aluminum substrate, we believe a higher or larger post (that does not protrude the droplet) can provide more liquid-solid interface area with only a minor increase in the thermal resistance. However, for a material with lower conductivity, the higher or larger post structure can provide extra interface area but also increase the thermal resistance. In that case, an optimal set of post parameters should exist – determining them is beyond the scope of this study, though.

# 7 Bubble generation and boiling behavior

In the main article, we explained the general improvement on cooling capacity of the post substrate based on an analytical heat transfer analysis. However, the temperature history and the droplet lifetime are very similar for substrate temperatures of 150°C, which cannot be explained by the theoretical analysis. Hence, we investigated the bubble generation and boiling behavior using high-speed imaging, shown in Fig. S8.

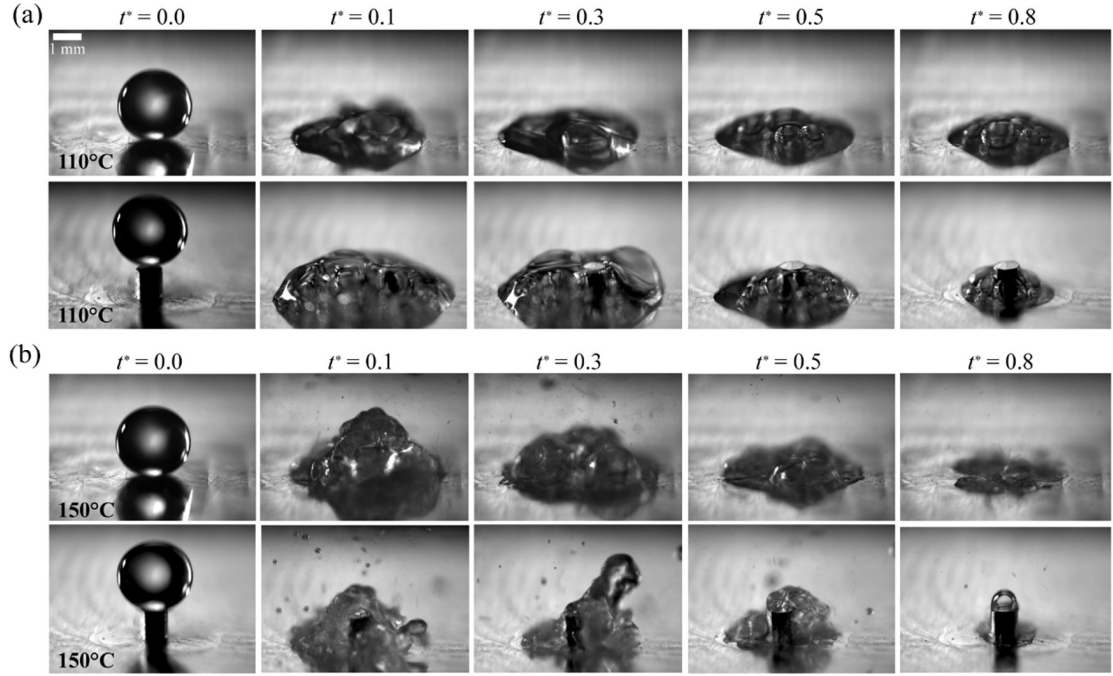

Fig. S8 High-speed snapshots of the bubble generation and boiling behavior at (a) $T_S = 110°C$ and (b) $T_S = 150°C$. The view is tilted from the horizontal by approximately 5° for better visualization of the post structure and the bubbles inside the droplet.

At $T_S = 110°C$ (Fig. S8 (a)), the droplet on the flat substrate experiences a gentle nucleate boiling, in which the generation and coalescence of small vapor bubbles lead to a slight deformation of the droplet shape but no significant lateral expansion. On the contrary, the droplet on the post substrate generates larger vapor bubbles and undergoes a foaming process as the bubbles accumulate and attach to the post, which ultimately leads to an expansion of the droplet-solid contact line ($t^* = 0.1$). The physical mechanism behind this foaming is not yet fully understood [2]. It has been suggested that small amounts of dissolved ionic salt in the water prevent bubble coalescence due to electrostatic effects and changes in surface tension [3]. In the present case, the post structure provides additional nucleation sites for the bubbles that can pin to the side wall of the post as they are growing ($t^* = 0.1$ and $t^* = 0.3$). As a result, the vapor pushes the liquid-solid contact line away from the post, leading to a much larger apparent liquid-solid interface area compared to the flat substrate, where foaming is absent. This increase in the contact area enhances the heat transfer, leading to the observed shorter droplet lifetime (Fig. 2 of main article) and lower surface temperature (Fig. 7 of main article) on the post substrate.

At $T_S = 150°C$ (Fig. S8 (b)), the boiling is vigorous and chaotic on both substrates. The droplet surface is deformed heavily by the intense vapor generation and the atomization near the liquid rim.

Except for a slightly more centered droplet on the post substrate, the boiling behavior for the two substrates is quite similar. Due to the distortion of the droplet, the post structure is partially exposed, reducing the influence of the (nominally) additional liquid-solid contact area compared to the flat substrate. This observation explains the minimal difference in the substrate temperatures shown in Fig. 7 (c) of the main article.

# 8 Cooling capacity of multi-droplet impact

We presented and discussed the cooling capacity of single-droplet impact in the main manuscript and extend our exploration to multi-droplet impact here. Fig. S9 shows the area-weighted average temperature of the bottom surface temperature at different $T_S$. A strong temperature drop can be identified once a new droplet impacts the substrate, followed by a recovery as the droplet evaporates. However, if the volume flow rate is much larger than the evaporation rate, the droplets can grow to a very large body of water (a water pool). Then, for these flooded samples, we can expect that both the temperature drop and its recovery will be smaller due to the slow temperature change of the water pool. For the example shown at $T_S = 110°C$, where the volume flow rate was low enough to prevent flooding, the temperature of the post substrate was always lower than that of the flat substrate, indicating that the post structure continuously increases the heat transfer in this temperature range and enables an efficient increase in cooling capacity. For $T_S = 130°C$, we can identify a higher temperature drop on the post substrate for the first two droplet impact events. After that, the sample flooded due to a higher volume flow rate compared to the 110°C-case. The flooding minimized the influence of the post and the temperature curves on the two substrates gradually became similar. At $T_S = 150°C$, the heat supply to the droplet was sufficient to prevent flooding, and – similar to the 110°C-case – the post substrate showed sustained larger temperature drops than the flat substrate. For $T_S = 165°C$, due to the Leidenfrost effect on the flat surface, the post substrate showed a much lower temperature, as expected. These results demonstrate that the post substrate can promote a higher cooling capacity for multi-droplet impact for as long as the substrate is not flooded.

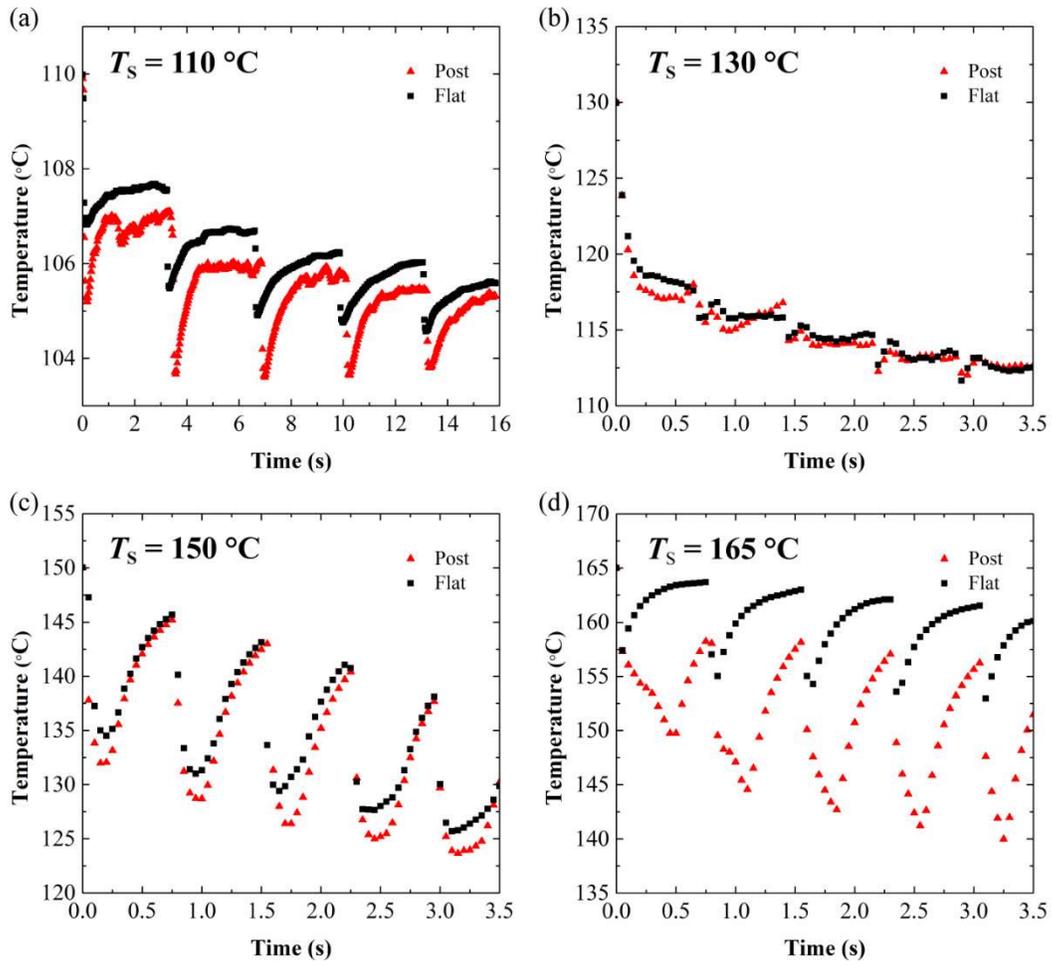

Fig. S9 Area-weighted average bottom surface temperature for multi-droplet impact at (a) $T_S$ = 110°C, (b) $T_S$ = 130°C, (c) $T_S$ = 150°C, and (d) $T_S$ = 165°C. The volume flow rate is 31.5 μL/s for 110°C, and 137.5 μL/s for all other cases. For all the experiments $We \approx 20$.

# 9 Droplet impact on inclined surfaces

Fig. S10 provides the droplet dynamics and the backside temperature distribution for droplet impact at the initial temperature of $T_S = 160°C$. Similar to the case of 145°C from Fig. 8 in the main article, on the flat substrate, the droplet again slides down the substrate ($t = 10.0$ ms), but then directly bounces off the surface ($t = 25.0$ ms). The back-side temperature briefly decrease around the impact region, followed by a quick recovery after the droplet bounces off the surface ($t = 50.0$ ms). The flat substrate fails to maintain a resonable cooling performance. On the post substrate, the droplet attaches to the left side of the post due to the mixed effects of pining and gravity ($t = 10.0$ ms). Compared to a symmetric shape at $T_S = 145°C$, this asymmetric shape indicates that the base surface almost enters the film boiling regime, where the droplet is levitated by the vapor layer and pins only to the post. Then the droplet remains attached to the left side of the post and quickly vaporizes ($t = 25.0$ ms, $t = 50.0$ ms). The corresponding temperature profile shows a well-defined low-temperature region, which indicates an effective cooling performance around the impact area.

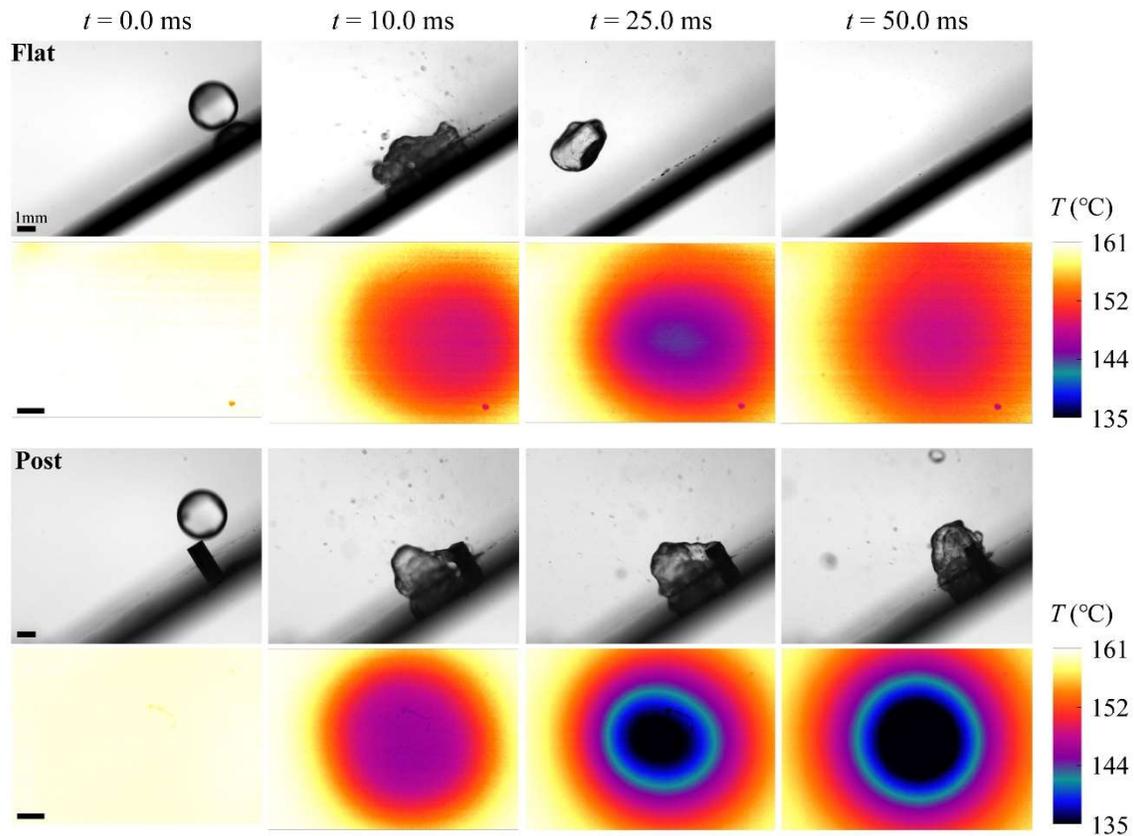

Fig. S10 High-speed and IR imaging sequence of droplets impacting on a flat and a post substrate inclined at 30° at an initial temperature of $T_S = 160°C$.

We also investigated the droplet dynamics for off-center droplet impact on the post substrate. For an impact location uphill from the post, we can expect that the droplet will be trapped by the post structure as it is sliding down toward the post. For an impact location downhill from the post, it is easily conceivable that the post can only pin the droplet if there is contact, *i.e.*, if the droplet is able to touch the post upon spreading. Hence we tested the droplet impact with downhill impact locations, shown in Fig. S11. The off-center distance $\Delta x$ is measured as the horizontal distance from impact location to the center of the post (on the base surface). We found that the post structure can still prevent the sliding and bouncing of the droplet for $\Delta x \leq 1.6$ mm downhill (recall, droplet diameter $D = 2.7$ mm). Compared to a center impact at the same substrate temperature (Fig. 8 in the main article), the droplet during off-center impact is initially more deformed, with an uphill "tail" towards the post ($t = 20.0$ ms). However, eventually the pinning force is able to pull the bulk of the droplet back towards the post ($t = 40.0$ ms), where the droplet remains attached to the post structure and evaporates ($t = 80.0$ ms).

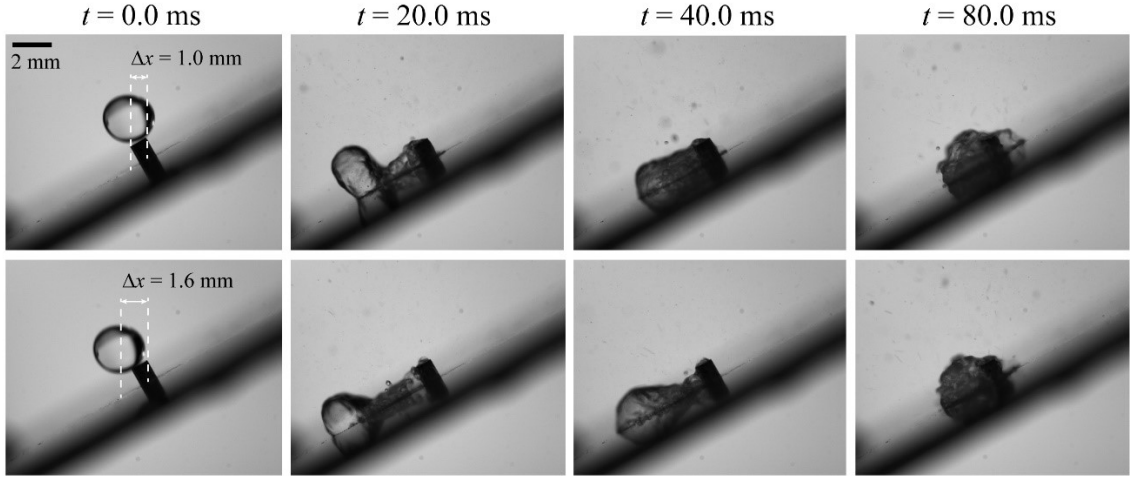

Fig. S11 High-speed imaging sequence of off-center droplets impacting the post substrates inclined at 30° at initial temperatures of $T_S = 145$°C. The horizontal distance from the impact location to the center of the post is 1.0 mm (top) and 1.6 mm (bottom).